\documentclass[prd,aps,preprint,amsmath,nofootinbib,amssymb,superscriptaddress,floatfix]{revtex4-1}
\usepackage{graphicx}
\usepackage{subfigure}
\usepackage{hyperref}
\usepackage{array}
\usepackage{booktabs}
\usepackage{makecell}
\usepackage{multirow}
\usepackage{enumerate}

\begin{document}

%\title{Bath deformations and Page Curves in Holographic Superconductor}
\title{Page Curves in Holographic Superconductors}

\author{Yuanceng Xu}
\email{xyc@mails.ccnu.edu.cn}
\thanks{Corresponding author}
\affiliation{Institute of Astrophysics, Central China Normal University, Wuhan, Hubei 430079, China}

\author{Dong Wang}
\email{indulge_chen@163.com}
\affiliation{School of Sciences, Hunan University of Technology, Zhuzhou, Hunan 412008, China}

\author{Qiyuan Pan}
\email{panqiyuan@hunnu.edu.cn}
\affiliation{Key Laboratory of Low Dimensional Quantum Structures and Quantum Control of Ministry of Education, 
Synergetic Innovation Center for Quantum Effects and Applications, and Department of Physics, Hunan Normal University, Changsha, Hunan 410081, China}

\date{\today}

\begin{abstract}
Considering a doubly holographic model, we study the black hole information paradox for the eternal AdS$_{d}$-RN black hole 
coupled to and in equilibrium with a $d$-dimensional conformal bath whose
state has been deformed by the charged scalar field coupled to a U(1) gauge field. 
Without a brane, the spontaneous symmetry breaking of the gauge field on boundary systems 
can induce a second-order phase transition of the charged scalar field at the critical temperature, 
known as holographic superconductors.
The bath deformation can significantly change its entanglement dynamics with the black hole, 
resulting in variations in the Page curve and Page time.
Our results indicate that characteristic parameters of the Page curve, such as entanglement velocity,
initial area difference and Page time, can be used as suitable probes to detect superconducting phase transitions. 
In particular, the entanglement velocity can also probe both Kasner flows and Josephson oscillations.
When keeping the endpoint of the radiation region fixed at twice the critical Page point,
%the Page time monotonically increases with increasing deformation for the conformal dimension 
%$\Delta_{-}=1$ (corresponding to the ``Neumann'' quantization).
%However, for $\Delta_{+}=2$ (corresponding to the ``Dirichlet'' quantization),
%and charges $q=2, 3, 6,$ and $12,$ the Page time initially decreases and then increases with increasing deformation.
%For the same conformal dimension and fixed
%bath deformation, a larger charge leads to a smaller Page time.
%Furthermore, 
the entanglement velocity (the internal backreaction) 
has a more significant influence on the Page time compared to the initial area difference (the external backreaction).

\end{abstract}
\maketitle

\section{Introduction}

A recent study on the black hole information paradox confirms that informations within a black hole 
can be extracted through Hawking radiation, 
preserving the information conservation and the quantum-mechanical unitarity
 \cite{Penington:2019npb,Almheiri:2019psf,Almheiri:2019yqk,Almheiri:2019hni} (see e.g. \cite{Almheiri:2020cfm} for a review). 
Unlike previous calculations by Hawking \cite{Hawking:1976ra}, 
the novel research approach focuses on the computation of the fine-grained entropy of the late-time `Radiation' $\mathcal{R}$ region,
emphasizing the need to simultaneously consider the fine-grained entropy of an interior region within the black hole, 
referred to as the `Island' $\mathcal{I}$ whose boundary corresponds to a `Quantum Extreme Surface' (QES)\cite{Engelhardt:2014gca} $\partial \mathcal{I}$. 
\begin{eqnarray}
    S(R)=min \left\{ext  \left[    \frac{ Area(\partial \mathcal{I}) }{4 G_{N}} + S_{matter}(\mathcal{I}\cup \mathcal{R})     \right]    \right\}.
\end{eqnarray}
Due to the entanglement between antiparticles within the island and outward-radiating positive particles,
when considering the entanglement entropy of both regions together, particle-antiparticle pairs cancel each other out,
resulting in a total entanglement entropy smaller than that of the isolated radiation region. 
This alignment with the Page curve \cite{Page:1993wv,Page:2013dx} demonstrates that the Hawking radiation maintains unitarity, 
affirming the conservation of the black hole information. 

Early investigations into the island rule primarily centered around two-dimensional dilaton gravity models.
Because the two-dimensional gravity allows for replica wormholes \cite{Penington:2019kki,Almheiri:2019qdq},
enabling a more rigorous derivation of the island formula. Subsequently, the island rule was extended to higher-dimensional scenarios 
through the use of doubly-holographic models \cite{Almheiri:2019hni,Almheiri:2019psy}:
the $d$-dimensional anti-de Sitter (AdS$_d$) gravity theory coupled 
to the matter (conformal field theory (CFT$_d$)) that has a ($d+1$)-dimensional holographic dual, 
which is a higher-dimensional version of the Randall-Sundrum (RS) setup \cite{Randall:1999vf,Randall:1999ee}.
The AdS$_d$-Einstein gravity coupled with the CFT$_d$-matter is referred to as the RS brane or Planck brane.
It is termed ``doubly-holography'' because it leverages two instances of the AdS/CFT holographic duality, 
thus establishing connections among three equivalent descriptions for the same system
\cite{Karch:2000ct,Karch:2000gx,Geng:2020qvw,Chandrasekaran:2020qtn,Krishnan:2020fer,Chen:2020jvn,Ling:2020laa,Ling:2021vxe,Geng:2020fxl,Bhattacharya:2021jrn,
Ghosh:2021axl,Geng:2021iyq,Uhlemann:2021nhu,Neuenfeld:2021bsb,Caceres:2021fuw,Omiya:2021olc,Geng:2021hlu,Sun:2021dfl,Chou:2021boq,Geng:2021mic,
Hu:2022ymx,Anous:2022wqh,Demulder:2022aij,Basu:2022crn,Karch:2023ekf,Jeong:2023hrb}:

\begin{enumerate}
   
\item[(\uppercase\expandafter{\romannumeral1})] \textbf{Boundary Perspective:} A $d$-dimensional boundary conformal field theory (BCFT$_d$) %\cite{Cardy:2004hm,McAvity:1995zd}) 
with a ($d-1$)-dimensional boundary. %\cite{Takayanagi:2011zk,Fujita:2011fp}

\item[(\uppercase\expandafter{\romannumeral2})]  \textbf{Bulk Perspective:} An asymptotically AdS$_{d+1}$ Einstein gravity with an AdS$_d$ RS brane.

%Karch-Randall(KR) braneworlds \cite{Karch:2000ct,Karch:2000gx,Karch:2020iit} %or $critical$ tension Randall-Sundrum(RS) braneworlds 
%\cite{Randall:1999vf,Randall:1999ee} as an ``end-of-the-world"(EOW) brane

\item[(\uppercase\expandafter{\romannumeral3})]  \textbf{Brane Perspective:} A CFT$_d$ living on the AdS$_d$ brane with a boundary CFT$_d$ bath on the flat space 
linked by transparent boundary conditions.

\end{enumerate}

\begin{figure}[htb]
	\subfigure[]{
	\includegraphics[width=0.35\textwidth,height=0.17\textwidth]{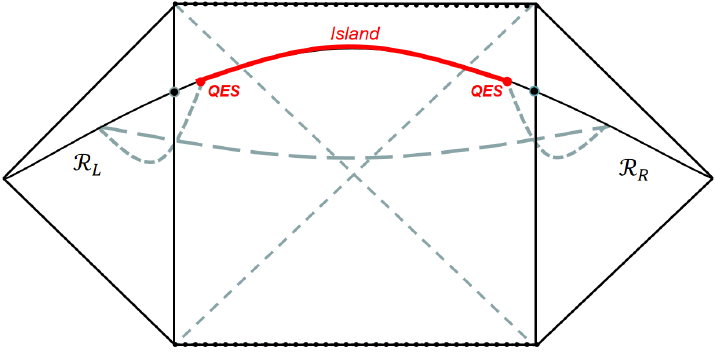}}
    \hfill
	\subfigure[]{
	\includegraphics[width=0.3\textwidth,height=0.25\textwidth]{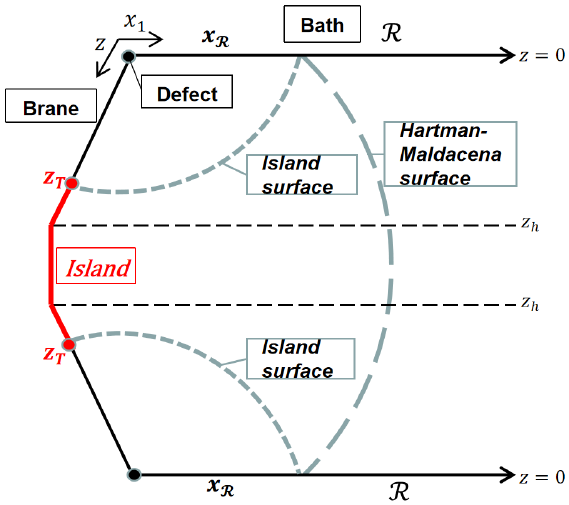}}
    \hfill
    \subfigure[]{
    \includegraphics[width=0.3\textwidth,height=0.2\textwidth]{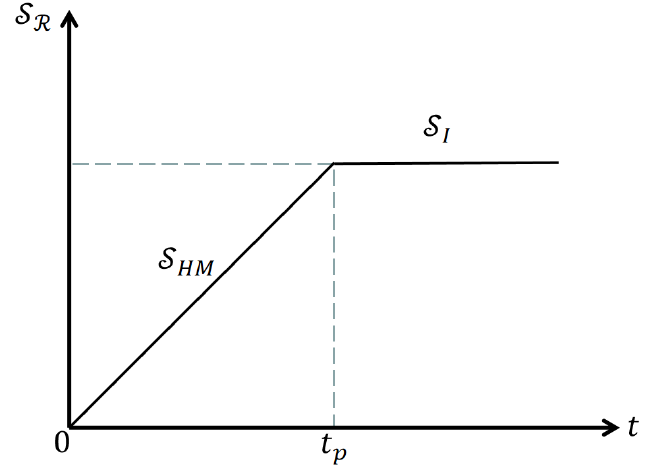}}
	\caption{$(a)$ The eternal two-side black hole and the island which emerges after Page time with QES outside the horizon.
    $(b)$ Two different candidate RT minimal surfaces in a doubly holographic setup.
    $(c)$ The early part of the Page curve is determined by the time-dependent HM surface, 
    while the later part of the Page curve is determined by the constant island surface.} \label{FIG00}
\end{figure}

The description (\uppercase\expandafter{\romannumeral3}) precisely simulates the Hawking radiation process of a black hole. 
The advantage of doubly holography is that the entanglement dynamics between 
the black hole and radiation region are reflected instantaneously 
in the corresponding entanglement wedges (EW) in the bulk of description (\uppercase\expandafter{\romannumeral2}). 
Then, according to the Ryu-Takayanagi/Hubeny-Rangamani-Takayanagi (RT/HRT) formula \cite{Ryu:2006bv,Hubeny:2007xt}, 
by replacing the computation of QES's areas with the computation of RT/HRT surface's areas 
in the bulk of the description (\uppercase\expandafter{\romannumeral2}) at different times, 
we can obtain the entanglement entropy of the CFT in the black hole and radiation region at different stages of the radiation.
Specifically, we consider a two-sided black hole model involving two black holes and two baths.
More precisely, we would like to fix one end of the RT surface at the conformal boundary at a given location $x_{\mathcal{R}}$. 
As we extend from the $x_{\mathcal{R}}$ of the boundary into the bulk, 
two candidate RT surfaces of interest. 
One is the extremal surface with the endpoint terminating on the brane,
called the ``island surface". 
The other is the extremal surface that passes through the black hole event horizon, 
reaches a radial critical point $z_{*}$ inside the black hole, 
and then crosses out of the horizon to reach another boundary,
the ``Hartman-Maldacena (HM) surface" \cite{Hartman:2013qma}. 
Two candidate extremal surfaces in the doubly holographic model are presented in FIG. \ref{FIG00}(b).
Because the Einstein-Rosen (ER) bridge grows with time, 
the area of the HM surfaces increases from the initial time ($t=0$). 
On the other hand, the area of the island surface, is time-independent, making it a constant. 
FIG. \ref{FIG00}(c) is the time-evolution curve of the radiation region's entanglement entropy, known as the Page curve.
The initial part of the Page curve, which increases with time, is obtained by calculating the area of time-dependent HM surfaces. 
The latter part of the Page curve, independent of time, is obtained by calculating the area of island surfaces.
In order to obtain a reasonable Page curve, the area of the HM surface at the initial time ($t=0$) must be smaller than 
the area of the island surface. 
For this, the endpoint $x_{\mathcal{R}}$ of the radiation region must be larger than the ``Page point" $x_{p}$,
which is the value of $x_{\mathcal{R}}$ when the initial HM surface's area is equal to the island surface's area.
Both candidate extremal surfaces are divergent at the UV conformal boundary, so the standard holographic renormalization will be used to remove
these divergent terms by a cutoff ($\epsilon=z/z_{h} \ll 1$) at the boundary.

The description (\uppercase\expandafter{\romannumeral3}) primarily discusses the coupling of a RS brane 
and a non-gravitating bath on the conformal boundary with the imposition of transport boundary conditions at their intersection (i.e., the defect). 
Due to conformal baths and transport boundary conditions, the gravity on the brane becomes massive \cite{Geng:2020qvw}, 
which raises a fundamental question: is it necessary, in our physical reality, for gravitons to have mass to address the black hole information paradox? 
In other words, do we require such a bath for a comprehensive understanding of our universe?
In fact, the bath gives too many computational constraints. 
To investigate to what extent the presence of the bath affects the explanation of the black hole information paradox, 
the author of the paper \cite{Caceres:2021fuw} introduces a scalar operator $\phi $ as a new,
tunable scale to deform the bath and break the conformal invariance,
which allows us to investigate how different degrees of bath deformation influence the Page curve.
The deformation triggers a holographic RG flow \cite{Balasubramanian:1999jd,deBoer:2000cz,Bianchi:2001kw,Fukuma:2002sb}.
One of the most significant conclusions in \cite{Caceres:2021fuw} is that the Page curve can serve as a probe for the holographic RG flow.
We hope to extend the research presented in paper \cite{Caceres:2021fuw} further. 
Specifically, we intend to deform the bath using a charged scalar field, 
which involves the coupling of the charged scalar field with a U(1) gauge field.
The U(1) symmetry is broken by the gauge field, inducing a second-order phase transition of the charged scalar field at the critical temperature,
known as holographic superconductors
\cite{Gubser:2008px,Hartnoll:2008vx,Hartnoll:2008kx,Hartnoll:2009sz,Herzog:2009xv,Horowitz:2010gk,Horowitz:2008bn,Cai:2015cya}. 
The primary motivation behind this research is to explore, 
within the context of the doubly holographic setup, 
not only whether the Page curve can probe the Kasner flows but, 
more importantly, whether the Page curve and Page time can track the informations of the superconducting phase transition.
Similar to the neutral case \cite{Frenkel:2020ysx,Wang:2020nkd}, with no branes, 
introducing a charged scalar field on the conformal boundary transforms the geometry near the singularity of the 
AdS-Reissner-Nordstr$\ddot{o}$m (AdS-RN) black hole into a Kasner universe.
Furthermore, it is noteworthy that the charged scalar hair brings about significant changes in the dynamical structure inside the AdS-RN black hole.
Specifically, the black hole's interior dynamics are cleanly separated into three different epochs \cite{Hartnoll:2020rwq,Hartnoll:2020fhc}:
the collapse of ER bridge, Joseph oscillations and Kasner universe; in some cases, it even undergoes a process known as Kasner inversions. 
Eventually, spacetime terminates at a spacelike Kasner singularity,
and it can be proven that there is no Cauchy horizon \cite{Cai:2020wrp,Sword:2021pfm,Grandi:2021ajl,An:2021plu,Devecioglu:2023hmn}.

The rest of the paper is organized as follows. 
In section \ref{section02}, 
we begin by constructing the necessary doubly holographic model, 
ensuring that the geometry of the tensionless brane and all matter fields satisfy the Neumann boundary conditions.
In section \ref{section03}, we review the holographic superconductors with backreaction. 
In section \ref{section04}, we review the internal structure of the black hole in the holographic superconductivity model.
In section \ref{section05}, we get the Page curve under bath deformation by calculating the area of the HM surface and the island surface.  
In final section \ref{section06}, we present the conclusions drawn from our study and the work we are interested in for the future.

\section{The doubly-holographic setup}  \label{section02}

The entire action of the $(d+1)$-dimensional asymptotic AdS$_{d+1}$ spacetime considering a doubly holographic setup reads
\begin{eqnarray}
    I=\frac{1}{16\pi G^{(d+1)}_{N}}&& \left[ \int_{M} d^{d+1}x\sqrt{-g }\left( R-2\Lambda+\mathcal{L}_{m}  \right)+
    2\int_{B} d^{d}x\sqrt{-h}\left( K-\alpha\right) \right. \nonumber \\
    &&\left. +2\int_{\partial}d^{d}x\sqrt{-h_{\partial}}K_{\partial}-2\int_{B\cap \partial }d^{d-1}x\sqrt{-\Sigma } \theta \right],  \label{fullaction}
\end{eqnarray}
where $\mathcal{L}_{m}$ denotes the Lagrange density of all matter fields in the bulk. 
Here, $B$ is the RS brane with tension $\alpha $, 
which serves as a gravitational part of the boundary of the bulk spacetime. $K=h^{ab}K_{ab}$ is the trace of the extrinsic curvature $K_{ab}$ defined by 
$K_{ab}=h_{a}^{c}\triangledown_{c}n_{b}$, where $n$ represents the unit normal vector on the brane. 
The first term in the second line of the above action (\ref{fullaction}) is the Gibbons-Hawking boundary term, where $K_{\partial}$ 
is the trace of the extrinsic curvature on the conformal boundary $\partial$ (as the non-gravitational bath).
The last term represents the junction term at the intersection of the brane and conformal bath.
$\theta$ is the angle between the brane and conformal bath. $g$, $h$, $h_\partial$, and $\Sigma$ 
denote the matrix determinant of $M$, $B$, $\partial$, and $B\cap\partial$, respectively.
Taking a variation with respect to the metric in the action (\ref{fullaction}), yields the Einstein field equation
\begin{eqnarray}
    G_{\mu \nu }+\Lambda g_{\mu \nu }=\kappa T_{\mu \nu }.
\end{eqnarray}
In this paper, lowercase Greek letters ($\mu, \nu $) represent ($d+1$) dimensions,
while lowercase Latin letters ($a,b$) represent $d$ dimensions.
When we impose the following Neumann boundary condition on the 
gravitational part of the boundary, the RS brane is called the Planck brane
\begin{eqnarray}
    K_{ab}-Kh_{ab}+\alpha h_{ab}=0   \label{neumann}.
\end{eqnarray}
We will start with the generic asymptotically AdS$_{d+1}$ planar black hole written in Poincar$\acute{e}$ coordinates
\begin{eqnarray}
    ds^2=\frac{1}{z^2}\left(-fe^{-\chi}dt^2+\frac{dz^2}{f}+dx^{2}_{1}+\sum_{i = 2}^{d-2}dx^{2}_{i}     \right).
\end{eqnarray} 

\begin{figure}[htb]
	\subfigure[]{
	\includegraphics[width=0.6\textwidth,height=0.23\textwidth]{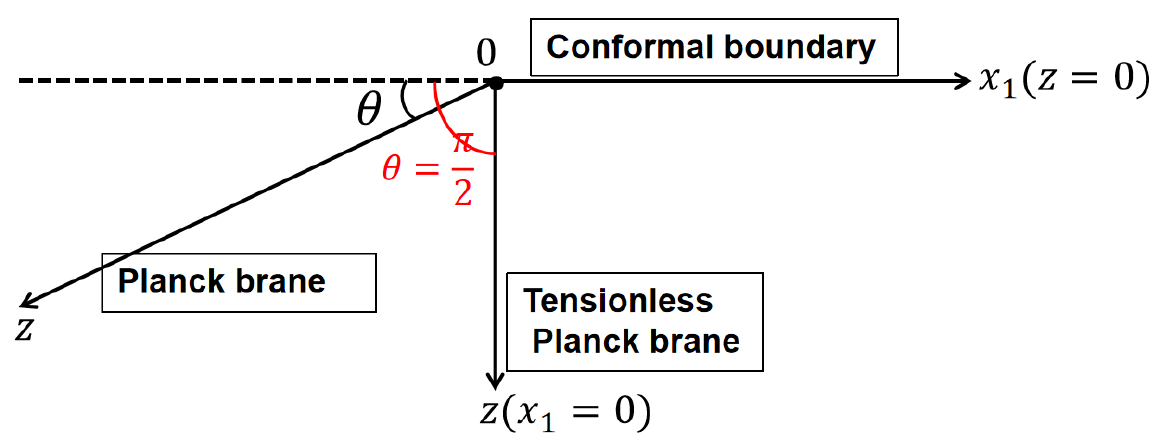}}
    \hfill
	\caption{A schematic diagram of the Planck brane and the tensionless Planck brane for the tension $\alpha=0$ with $\theta=\pi/2$.} \label{FIG01}
\end{figure} 
Now, we will consider the surface $z=-x_{1}\tan \theta $ as the Planck brane, 
where $\theta $ is the angle between the brane and the conformal boundary, as see in FIG. \ref{FIG01}.
After computing the extrinsic curvature of the surface
\begin{eqnarray}
    K_{ab}=h_{ab}\cos \theta     \label{extcur},
\end{eqnarray}
we can get the tension by using the Neumann boundary condition (\ref{neumann})
\begin{eqnarray}
    \alpha=(d-1)\cos \theta .
\end{eqnarray}
When the tension satisfies the following bound, the RS brane is subcritical tension brane,
also known as the Karch-Randall (KR) brane
\begin{eqnarray}
    |\alpha|<d-1.
\end{eqnarray}
With no scalar field and U(1) gauge field, the induced geometry of the KR brane is pure AdS$_{d}$ spacetime.
However, in this paper, we will consider the bath deformation of a charged scalar field. 
As a result, the brane is no longer a vacuum solution but a hair black hole. 
Nevertheless, as long as we find the boundary that satisfy Neumann boundary condition (\ref{neumann}), 
the non-vacuum solution can still be a KR brane.
For simplicity in our calculations, we only consider the case where the tension is absent 
\begin{eqnarray}
    \alpha=0 \   \  (i.e., \theta=\frac{\pi }{2}). 
\end{eqnarray}
In this setup, the brane has no backreaction on the background, 
but what we concern is the backreaction of the matter field.
Due to $\theta=\pi/2$ corresponding to the tensionless limit, 
for convenience, we can take the hypersurface
\begin{eqnarray}
    x_{1}=0,
\end{eqnarray}
as the position of the brane, as shown in FIG. \ref{FIG01}.
We must ensure that this hypersurface is indeed the appropriate KR brane,
meaning that the background geometry, the charged scalar field, 
and the gauge field all satisfy Neumann boundary condition (\ref{neumann}) on the hypersurface.
The unit normal vector on the hypersurface is $n_{\mu }=\frac{\delta_{\mu 1}}{z}$,
then the extrinsic curvature is 
\begin{eqnarray}
    K_{ab}=h^{c}_{a}\triangledown_{c}n_{b}=0.
\end{eqnarray}
We can also find $K_{ab}=0$ from the result (\ref{extcur}) when $\theta=\pi/2$.
So this is indeed a tensionless KR brane.
For the U(1) gauge field $A_{\mu }$, 
we can impose the following Neumann boundary condition on the brane \cite{Chu:2018ntx,Miao:2018qkc,Ling:2020laa}
\begin{eqnarray}
    F_{\mu \nu }n^{\mu }h^{\nu }_{a}=0.
\end{eqnarray}
As for the charged scalar field ($\phi=\phi(z)$) being independent of the coordinate $x_{1}$, 
it evidently satisfies a Neumann boundary condition
\begin{eqnarray}
    n^{\mu}\partial_{\mu }\phi(z)=z\partial_{1}\phi (z)=0.
\end{eqnarray}

\section{Holographic Supercoductor model}   \label{section03}

\begin{figure}[htb]
    	\centering
    	\subfigure[]{
    	\includegraphics[width=0.475\textwidth,height=0.35\textwidth]{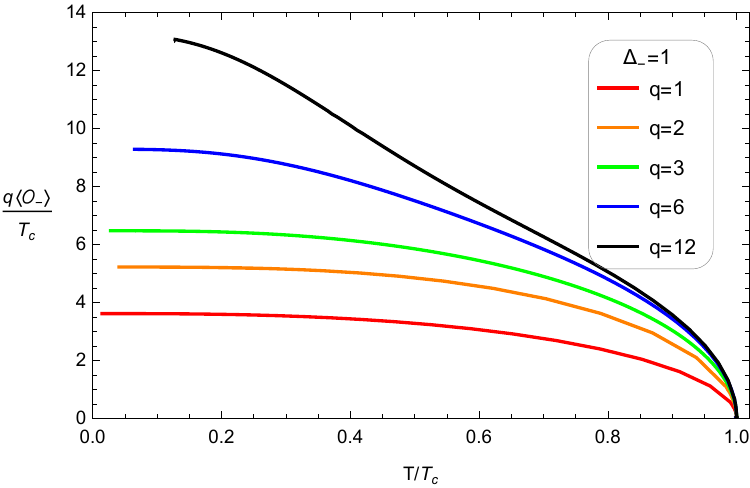}}
        \hfill
    	\subfigure[]{
    	\includegraphics[width=0.475\textwidth,height=0.35\textwidth]{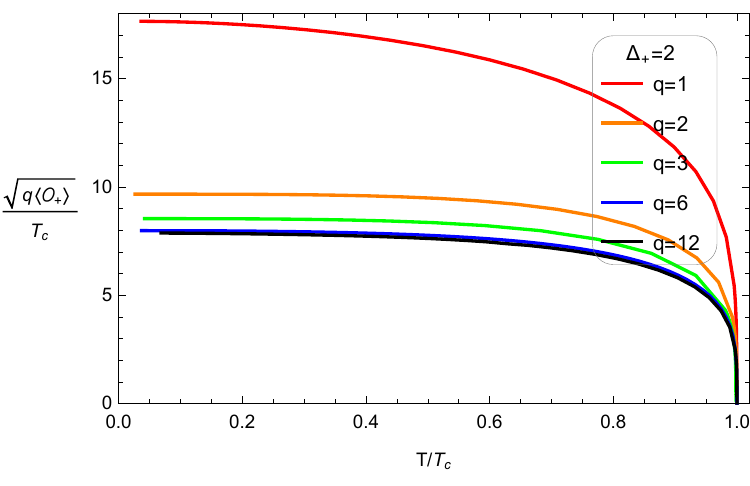}}
    	\caption{The value of the condensate as a function of the temperature, 
        with the fixed charge of the scalar field, for two operators $\mathcal{O}_{-}$ and $\mathcal{O}_{+}$ with conformal dimensions $\Delta_{-}$ and  $\Delta_{+}$, respectively.} \label{condensate2}
    \end{figure}

In this section, we provide a brief overview of the holographic superconductors with the backreaction. 
For a more in-depth understanding, readers are encouraged to refer to the references 
\cite{Gubser:2008px,Hartnoll:2008vx,Hartnoll:2008kx,Hartnoll:2009sz,Herzog:2009xv,Horowitz:2010gk,Horowitz:2008bn,Cai:2015cya}.
The action of the bulk is the ($d+1$)-dimensional AdS$_{d+1}$ Einstein-Maxwell theory with a scalar field coupling U(1) gauge field
\begin{eqnarray}
    I_{bulk}=\frac{1}{16\pi G^{(d+1)}_{N}}\int d^{d+1}x\sqrt{-g}\left[R-2\Lambda -\frac{1}{4}F^2-|D\phi|^2 -m^2|\phi|^2   \right],  
\end{eqnarray}
where $G^{(d+1)}_{N}$ is the $(d+1)$-dimensional Newtonian gravitational constant, and $R$ is the Ricci scalar of spacetime. The negative cosmological constant is
\begin{eqnarray}
    \Lambda =-\frac{d(d-1)}{2L^2},
\end{eqnarray}
where $L$ is the AdS radius and we will set $L=1$ from here. $F=dA $ is the field strength of the gauge field $A$.
$\phi $ is a complex scalar field with mass $m$. 
The covariant derivative is defined by $ D=\triangledown-iqA$,
where $q$ is the charge of the charged scalar field.
We will consider the following ansatz for the generic asymptotically AdS$_{d+1}$ planar black hole and matter fields
\begin{eqnarray} \label{met}
    ds^2=\frac{1}{z^2}\left(-fe^{-\chi}dt^2+\frac{dz^2}{f}+\sum_{i = 1}^{d-1}dx^{2}_{i}     \right) ,
\end{eqnarray}
\begin{eqnarray}
    \phi =\phi (z),\       \ A_{\mu }dx^{\mu }=A_{t}(z)dt.
\end{eqnarray}
The Hawking temperature is 
\begin{eqnarray}
    T=-\frac{f^{'}e^{-\chi/2}}{4\pi} \Bigg |_{z=z_{h}} .
\end{eqnarray}
We can get the equations of motion with four functions: $\phi (z)$, $A_{t}(z)$, $\chi(z)$, $f(z)$
\begin{eqnarray} 
&&\phi^{''}-\left(\frac{d-1}{z}-\frac{f^{'}}{f}+\frac{\chi^{'}}{2} \right)\phi^{'}+\left(\frac{q^2A^{2}_{t}e^{\chi}}{f^2}-\frac{m^2}{z^2f}  \right)\phi=0,  \label{eom1}\\ 
&&A^{''}_{t}-\left(\frac{d-3}{z}-\frac{\chi^{'}}{2}  \right)A^{'}_{t}-\frac{2q^2A_{t}\phi^2}{z^2f}=0,  \label{eom2}\\
&&\chi^{'}-\frac{2}{d-1}\left(z\phi^{'2}+\frac{zq^2A^{2}_{t}\phi^2}{f^2}e^{\chi } \right)=0,   \label{eom3}\\
%&&f^{'}+\frac{d}{z}-\left(\frac{d}{z}+\frac{z\phi^{'2}}{d-1} \right)f-\frac{1}{d-1}\left(\frac{zq^2A^{2}_{t}\phi^2e^{\chi}}{f}+\frac{z^3A^{'2}_{t}e^{\chi}}{2}+\frac{m^2\phi ^2}{z} \right)=0 \label{eom4} \\
&&f^{'}+\frac{d}{z}-\left(\frac{d}{z}+\frac{\chi^{'}}{2}  \right)f-\frac{1}{d-1}\left(\frac{z^3A^{'2}_{t}e^{\chi}}{2}+\frac{m^2\phi^2}{z} \right)=0,  \label{eom5}
\end{eqnarray}
where the prime denotes the derivative with respect to $z$.
For $\phi=0$, the analytical solution to the above equations of motion 
is the familiar $(d+1)$-dimensional AdS$_{d+1}$-RN black hole
\begin{eqnarray}
&&f(z)=1-\frac{z^d}{z^{d}_{h}}-\frac{d-2}{2(d-1)}\left(\frac{z^d}{z^{d}_{h}}-\frac{z^{2d-2}}{z^{2d-2}_{h}}   \right)z^{4}_{h}\rho ^2 ,\        \ \rho=\frac{\mu }{z_{h}},  \label{AdSRNBH} \\
&&A_{t}(z)=\mu\left(1-\frac{z^{d-2}}{z^{d-2}_{h}}\right),\       \   \chi(z)=0,   
\end{eqnarray}
where $z_{h}$ is the event horizon with $f(z_{h})=0$, $\mu$ is the chemical potential, 
and $\rho $ is the charge density.
The Hawking temperature for AdS$_{d+1}$-RN black hole is 
\begin{eqnarray}
    T_{RN}=\frac{1}{z_{h}}\left[\frac{d}{4\pi }-\frac{z^{4}_{h}(d-2)^2}{8\pi(d-1)}\rho^2 \right].
\end{eqnarray}
For $\phi\neq0$, we can expand the equations of motion (\ref{eom1})-(\ref{eom5}) 
near the horizon ($z\rightarrow z_{h}$) as considering $f(z_{h})=0$, $A_{t}(z_h)=0$
\begin{eqnarray}
    &&\phi(z)=\phi_{0}+\phi_{1}(z-z_h)+\phi_{2}(z-z_h)^2+\cdots , \\
    &&A_{t}(z)=A_{1}(z-z_{h})+A_{2}(z-z_{h})^2+\cdots,\\
    &&\chi (z)=\chi_{0}+\chi_{1}(z-z_{h})+\chi_{2}(z-z_{h})^2+\cdots,\\
    &&f(z)=f_{1}(z-z_{h})+f_{2}(z-z_{h})^2+\cdots.
\end{eqnarray}
Using the numerical shooting method, 
we can determine three constants $\phi_{0}$, $A_{1}$, and $\chi_{0}$ that yield solutions satisfying above equations of motion (\ref{eom1})-(\ref{eom5}) 
and the following boundary conditions.
Near the boundary $z\rightarrow 0$, the equations of motion (\ref{eom1})-(\ref{eom5}) have asymptotic solutions
\begin{eqnarray}
    \phi(z) \approx \phi_{-}z^{\Delta _{-}}+\phi_{+}z^{\Delta _{+}},\     \   A_{t}(z)\approx \mu -\rho z^{d-2},\     \ \chi(z)\approx 0,\       \ f(z)\approx 1, 
\end{eqnarray}
where the conformal dimension $\Delta _{\pm }$ is the dimension of the operator 
\begin{eqnarray}
    \Delta _{\pm }=\frac{d}{2}\pm \sqrt{\frac{d^2}{4}+m^2}. 
\end{eqnarray}
We can easily note that $\Delta=\Delta_{\pm }$ satisfy a mass-dimension renormalization
\begin{eqnarray}
    \Delta(\Delta-d)=m^2.
\end{eqnarray}
The AdS boundary is vacuum-stable only if $m^2\geq m^2_{BF}$. $m^2_{BF}=-d^2/4$ is the Breitenlohner-Freedman (BF) bound.
When $m^2=m^2_{BF}$, we have $\Delta_{-}=\Delta_{+}=d/2$. For $-d^2/4\leq m^2<-d^2/4+1$, 
there are two quantization schemes: according to the AdS/CFT dictionary,
$\phi_{+}$ is the vacuum expectation value $\left\langle \mathcal{O}_{+}  \right\rangle $, 
while $\phi_{-}$ is regarded as the source of the dual operator $\mathcal{O}$ and should vanish ($\phi_{-}=0$),
or $\phi_{-}$ is the vacuum expectation value $\left\langle \mathcal{O}_{-}  \right\rangle $, 
while $\phi_{+}$ is regarded as the source of the dual operator $\mathcal{O}$ and should vanish ($\phi_{+}=0$).
For $m^2\geq -d^2/4+1$, only the $\Delta_{+}$-mode can be normalized. 
In FIG. \ref{condensate2}, we provide the condensation of these two operators with different charges.
Note that we have fixed the dimension and the mass of the scalar field (i.e., $d=3$, $m^2=-2$). 
Unless otherwise stated, we maintain this setup in the following calculations.

\section{Kasner Solutions of Holographic Supercoductors}  \label{section04}

As mentioned in the introduction, 
the interior of the black hole with charged scalar hair exhibits intricate dynamical behaviors.
For instance, the relevant deformations can lead to the disappearance of the Cauchy horizon, 
triggering several distinct dynamical regimes within the black hole interior. 
These are the collapse of the ER bridge, Josephson oscillations, Kasner universe, and in some cases, 
a final Kasner inversion before reaching the Kasner singularity.
We will not delve into a detailed exposition of the complete picture.
Instead, we will provide a concise overview of the Kasner solutions.
Readers interested in a more in-depth understanding of the details are encouraged to refer to references 
\cite{Hartnoll:2020rwq,Hartnoll:2020fhc,Cai:2020wrp,Sword:2021pfm,Grandi:2021ajl,An:2021plu,Devecioglu:2023hmn}.

Near the singularity ($z\rightarrow \infty$), we assume that we can neglect the $A^{'}_{t}$ term in Eq. (\ref{eom5}),
the scalar-mass term and its charged term, then these equations of motion (\ref{eom1})-(\ref{eom5}) approximate into  
\begin{eqnarray}
   && \phi^{''}-\left(\frac{d-1}{z}-\frac{f^{'}}{f}+\frac{\chi^{'}}{2}   \right)\phi^{'}=0, \\
   && A^{''}_{t}-\left(\frac{d-3}{z}-\frac{\chi^{'}}{2}  \right)A^{'}_{t}=0, \\
   && \chi^{'}-\frac{2z\phi^{'2}}{d-1}=0,   \\
   && f^{'}-\left(\frac{d}{z}+\frac{\chi^{'}}{2}  \right)f=0.
\end{eqnarray}
The solutions of these equations are 
\begin{eqnarray}
    \phi=\alpha\sqrt{d-1}\log z,\  A^{'}_{t}=A_{0}z^{-3+d-\alpha^2},\        \ \chi=2\alpha^2\log z+\chi_{0},\        \  f=-f_{0}z^{d+\alpha^2},   \label{kasnersolution}
\end{eqnarray}
where $\alpha$, $A_{0}$, $\chi_{0}$, and $f_{0}$ are integration constants. By using the solutions (\ref{kasnersolution}), 
we can get metric components $g_{tt}$ and $g_{zz}$ 
\begin{eqnarray}
    g_{tt}=-\frac{fe^{-\chi }}{z^2}=f_{0}e^{-\chi_0}z^{-2+d-\alpha^2},\       \   g_{zz}=\frac{1}{z^2f(z)}=-\frac{z^{-2-d-\alpha^2}}{f_{0}}.
\end{eqnarray}

\begin{figure}[htb]
	\centering
	\subfigure[]{
	\includegraphics[width=0.7\textwidth,height=0.3\textwidth]{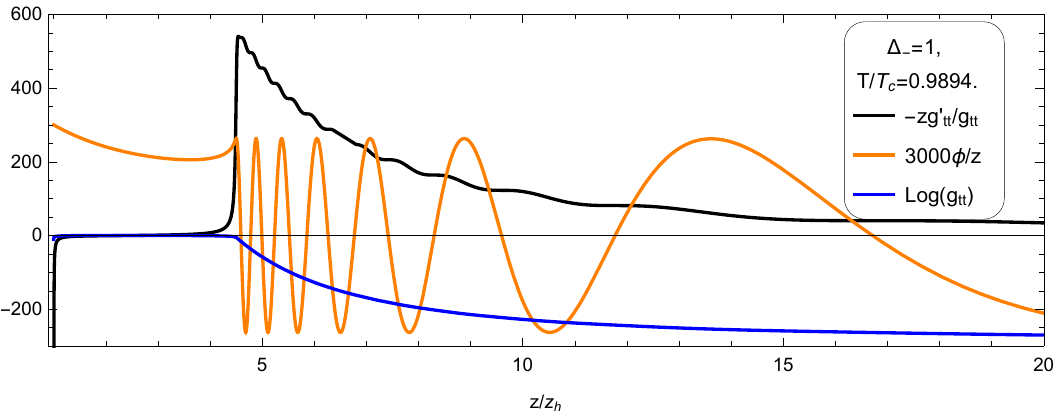}}
    \hfill\\
	\subfigure[]{
	\includegraphics[width=0.7\textwidth,height=0.3\textwidth]{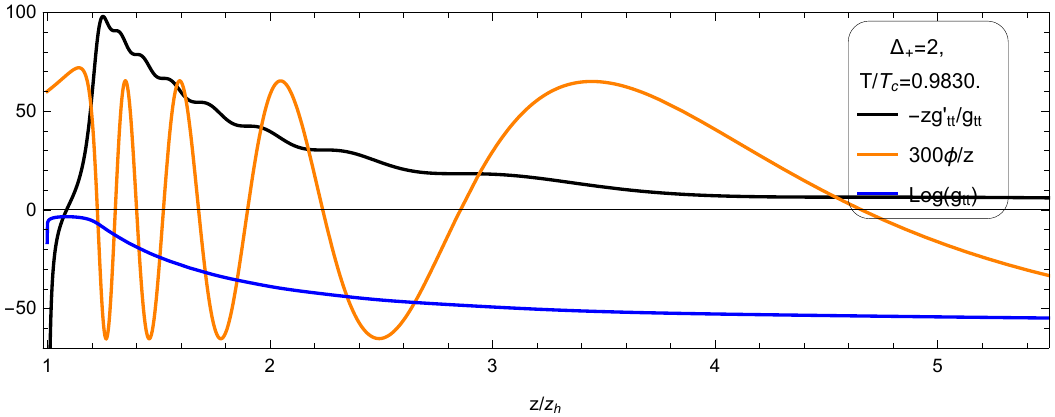}}
	\caption{$-z\frac{g'_{tt}}{g_{tt}}$, $3000\frac{\phi }{z}$, $300\frac{\phi }{z}$, $\log(g_{tt})$ are plotted 
    as a function of the radial coordinate $z/z_{h}$ for the two conformal dimensions and corresponding temperatures 
    (a): $\Delta_{-}=1$, $T/T_{c}=0.9894 $; (b): $\Delta_{+}=2$, $T/T_{c}=0.9830$. } \label{Josephson}
\end{figure}

We heve plotted $-zg'_{tt}/g_{tt}$, $\phi/z$, $\log(g_{tt})$ as a function of $z/z_{h}$ 
for two different conformal dimensions $\Delta_{\pm }$ in FIG. \ref{Josephson}.
We can find the distinct epochs of the collapse of the ER bridge and Josephson oscillations from the figure. 
The metric also captures the behaviors of the oscillations.
Consider the following parameter transformations
\begin{eqnarray}
    \tau = c_{\tau}z^{-\frac{d+\alpha ^2}{2}},\     \  \left( c_{\tau}=\frac{2}{(d+\alpha^2)\sqrt{f_{0}} }\right),
\end{eqnarray}
we can obtain the following Kasner solutions near the singularity of the spacetime \cite{kasner,kasner1,Belinski}
\begin{eqnarray}
    ds^2= -d\tau^2+c_{t}\tau ^{2p_{t}}dt^2+c_{x}\tau ^{2p_{x}}d\vec{x} ^2_{d-1},\      \   \phi \sim -p_{\phi } \log \tau,  
\end{eqnarray}
\begin{eqnarray}
    c_{t}=\frac{f_{0}e^{-\chi_0}}{c^{2p_{t}}_{\tau}},  \      \  c_{x}=\frac{1}{c^{2p_{x}}_{\tau }},
\end{eqnarray}
where the Kasner exponents $\{p_{t}$, $p_{x}$, $p_{\phi} \}$ are 
\begin{eqnarray}
    p_{t}=\frac{2-d+\alpha^2}{d+\alpha^2},\        \ p_{x}=\frac{2}{d+\alpha^2},\        \ p_{\phi}=\frac{2\sqrt{d-1}\alpha  }{d+\alpha^2},
\end{eqnarray}
which satisfy the Kasner constraints
\begin{eqnarray}
    p_{t}+(d-1)p_{x}=1,\       \ p^2_{t}+(d-1)p^2_{x}+p^{2}_{\phi}=1.
\end{eqnarray}

\begin{figure}[htb]
	\centering
	\subfigure[]{
	\includegraphics[width=0.475\textwidth,height=0.35\textwidth]{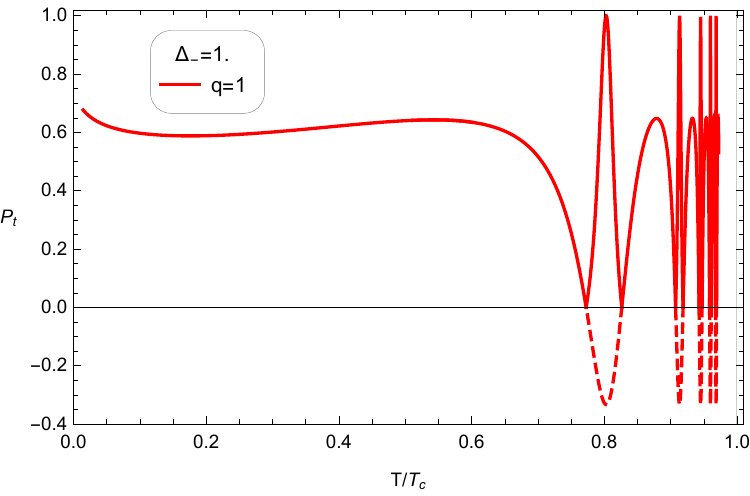}}s
    \hfill
	\subfigure[]{
	\includegraphics[width=0.475\textwidth,height=0.35\textwidth]{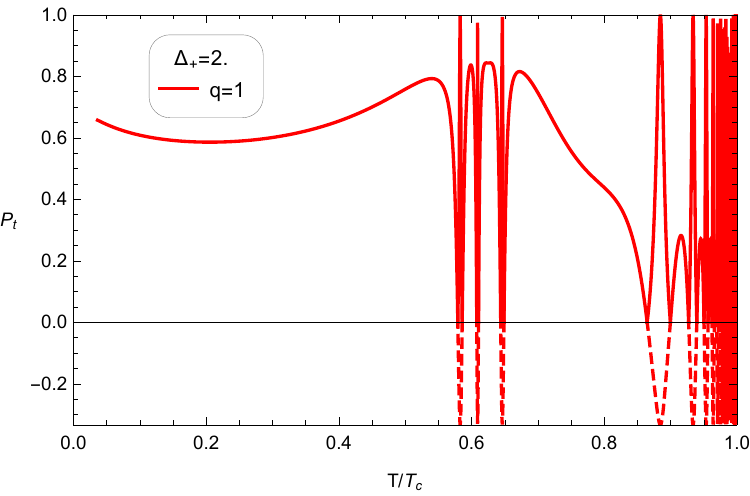}}
	\caption{The Kasner exponent $p_{t}$ as a function of the temperature $T/T_{c}$ after inversions. We have also shown 
     $p_{t}<0$ before inversions as a dashed red line. } \label{kasner}
\end{figure}

When $p_{t}<0$, corresponding to an ER bridge grows, describe a `Kasner inversion' $\alpha \rightarrow 1/\alpha $ in which
\begin{eqnarray}
    p_{t}\rightarrow -\frac{p_{t}}{2p_{t}+1}.
\end{eqnarray}  
In this case, the Kasner exponent $p_{t}$ changes its sign in the inversions at late times.
The first discovery of this inversion was made by Hartnoll and his collaborators in \cite{Hartnoll:2020rwq,Hartnoll:2020fhc}.
In FIG. \ref{kasner}, we show the variations of $p_t$ with $T/T_c$ 
for two different conformal dimensions $\Delta_{\pm }$ as charge ($q=1$).
The strong oscillations are shown near the critical temperature $T_{c}$. 
These oscillations in the Kasner exponents also reflect the imprint of Josephson oscillations.

\section{Bath deformations amd Page Curves}   \label{section05}

This section constitutes the central component of our article. 
Building upon the inspiration from the paper \cite{Caceres:2021fuw}, 
we investigate the Page curves for the eternal AdS$_{d+1}$ charged black hole within the context of a doubly holographic model,
where the BCFT's states in the conformal bath are deformed by a charged scalar field. 
The specific setup of this model is as follows: 
initially, we couple the brane with the conformal bath 
and impose transport boundary condition at the interface between the brane and the conformal bath, 
bringing them to thermodynamic equilibrium. 
As the Hawking radiation reaches the Page time, an interior region called the ``island" forms within the black hole 
(sometime, the island is outside the hoizon \cite{Almheiri:2019yqk}), 
causing the computed entanglement entropy of the radiation region to no longer increase continuously 
but instead exhibit an initial rise followed by a plateau. 
The resulting Page curves for this setup are termed the version of the eternal black hole, as shown in FIG. \ref{FIG00}(c).

\subsection{Growth Area of the Hartman-Maldacena Surfaces}  \label{GrowthAreaHM}
Before Page time, there is no contribution from island region. 
The candidate extremal surface is the time-dependent HM surface \cite{Hartman:2013qma}.
As time progresses, the extremal surface gets progressively closer to a specific critical radius within the interior of the black hole. 
At late times, the area of the extremal surface increases linearly with time.
When it gets close to the critical radius, the growth velocity of the extremal surface tends towards a constant value, 
which we call the entanglement velocity \cite{Hartman:2013qma,Carmi:2017jqz,Frenkel:2020ysx}. 
Since the entanglement velocity is highly sensitive to the geometric properties of the black hole's interior 
but not to the near-singularity region, we can use it as a probe for exploring the interior of the black hole.
We will briefly explain this fact and apply it to our model.
First, considering a constant-$x_{1}=x_{\mathcal{R}}$ slice, the induced metric takes the following form
\begin{eqnarray} 
    ds^2\mid_{x_{1}=x_{\mathcal{R}}}=\frac{1}{z^2}\left[-f(z)e^{-\chi(z)}dt^2+\frac{dz^2}{f(z)}+\sum_{i=2}^{d-1}dx^{2}_{i}     \right].
\end{eqnarray}
For a time-dependent surface ($z=z(t)$), the area function is 
\begin{eqnarray}
    A=V_{d-2}\int \frac{dt}{z(t)^{d-1}}\sqrt{-f[z(t)]e^{-\chi[z(t)]}+\frac{\dot{z}(t)^2}{f[z(t)]} },\  \ \dot{z}=\frac{dz}{dt},  
\end{eqnarray}
where $V_{d-2}=\int \prod^{d-1}_{i=2}dx_{i} $ is an overall volume factor. 
Since the volume factor is a constant, for convenience,
we can define the HM area density (which we will also refer to as ``area") and its corresponding Lagrangian
\begin{eqnarray}
    \mathcal{A}_{HM}=\frac{A}{v_{d-2}}=\int \frac{dt}{z(t)^{d-1}}\sqrt{-f[z(t)]e^{-\chi[z(t)]}+\frac{\dot{z}(t)^2}{f[z(t)]} }=\int dt \mathcal{L}.  \label{HMsurface}
\end{eqnarray}
Due to the explicit time-independent Lagrangian, we can identify that the ``energy" $E$ of the minimal surface is a constant of motion
\begin{eqnarray}
    E=\dot{z}\frac{\partial \mathcal{L}}{\partial \dot{z} }-\mathcal{L}=\frac{f(z)e^{-\chi (z)}}{z^{d-1}\sqrt{-f(z)e^{-\chi (z)}+\frac{\dot{z}^2 }{f(z)}} }, 
\end{eqnarray}
and the minimal surface has the following expression
\begin{eqnarray}
    \dot{z}=\pm f(z)e^{-\chi (z)/2}\sqrt{1+\frac{f(z)e^{-\chi(z)}}{z^{2(d-1)}E^2}}  \label{minimalsurface}.
\end{eqnarray} 
If there exists a location inside a black hole with a radius of $z_{*}$ satisfying the condition ($\frac{dz}{dt}|_{z=z_{*}}=0$), 
we can readily derive the constant $E$
\begin{eqnarray}
    E^2=-\frac{f(z_{*})e^{-\chi(z_{*})}}{z^{2(d-1)}_{*}} \label{energy}.
\end{eqnarray} 
By substituting (\ref{minimalsurface}) into (\ref{HMsurface}), 
and taking $dt=\frac{dz}{\dot{z}}$ into account,
we can get the area expression of the HM surface 
\footnote{The factor of $2$ in front of the integral is due to the consideration of two side black hole.}
\begin{eqnarray}
    \mathcal{A}_{HM}(t_b)=2\int^{z_{*}}_{0}\frac{dz}{z^{d-1}\sqrt{f(z)+e^{\chi(z)}z^{2(d-1)}E^2} }.   \label{HMsurface2}
\end{eqnarray}
Considering the following integral
\begin{eqnarray}
    \int^{z_{*}}_{0}\frac{dz}{\dot{z} }=\int^{t_{*}}_{t_{b}}dt=t_{*}-t_{b},
\end{eqnarray}
and the symmetry geodesic argument at the turning point ($t_{*}(z=z_{*})=0$), we can get the boundary time 
\begin{eqnarray}
    t_{b}=-P\int^{z_{*}}_{0}\frac{sgn(E)e^{\chi/2}}{f(z)\sqrt{1+f(z)e^{-\chi (z)}/z^{2(d-1)}E^2  }   }, \label{boundarytime}
\end{eqnarray}
where $P$ and $sgn(E)$ denote the principal value and the sign of the ``Energy" $E$, respectively.
Using the area expression (\ref{HMsurface2}) and boundary time (\ref{boundarytime}), we can numerically compute the time-dependent area of the HM surface.
To get the late-time behavior ($t_{b}\rightarrow \infty $), let's first consider the following function
\begin{eqnarray}
    g(z)=-\frac{f(z)e^{-\chi(z)}}{z^{2(d-1)}}.   \label{gzfunction}
\end{eqnarray}
After substituting the Kasner solutions (\ref{kasnersolution}) into function $g(z)$ (\ref{gzfunction}), we can find
\begin{eqnarray}
    g(z_{h})=0,\      \  \lim_{z\rightarrow \infty }g(z)=\lim_{z\rightarrow \infty }f_{0}e^{-\chi_{0}}z^{2-d-\alpha^2}=0,\         \ g(z)|_{z>z_{h}}>0.        
\end{eqnarray}
So, there is a critical radius $z_{c}$ inside the black hole that maximizes $g(z)$, and the critical radius also satisfies (\ref{energy}) 
\begin{eqnarray}
    E_{c}^2+\frac{f(z_{c})e^{-\chi(z_{c})}}{z^{2(d-1)}_{c}}=0.
\end{eqnarray}
When $z_{*}\rightarrow r_{c}$, the boundary time diverges ($t_{b}\rightarrow \infty $).
Next, by performing the derivative for the area expression of HM surface $\mathcal{A}_{HM}$ (\ref{HMsurface2}) with boundary time $t_{b}$  (\ref{boundarytime}) while also taking $f(z_c)<0$ into account,
we can readily obtain the growth rate of the HM surface area with respect to the boundary time 
\begin{eqnarray}
    \frac{\partial \mathcal{A}_{HM}}{\partial t_{b}}=2\frac{-f(z_{c})e^{-\chi(z_{c})}}{z^{2(d-1)}_{c}|E_c|}=2\sqrt{-\frac{f(z_c)e^{-\chi(z_c)}}{z^{2(d-1)}_{c}}}=\frac{2v}{z^{d-1}_{h}} ,  \label{envelo}
\end{eqnarray}
where $v$ is the entanglement velocity. For the AdS$_{d+1}$-Schwarzschild brane, the entanglement velocity is 
\begin{eqnarray}
     v_{AS}=\frac{\sqrt{d}(d-2)^{\frac{d-2}{2d}} }{[2(d-1)]^{\frac{d-1}{d}}}.
\end{eqnarray}
For the AdS$_{d+1}$-RN brane, the entanglement velocity is 
\begin{eqnarray}  \label{ARNvelocity}
    v_{ARN}=\sqrt{\frac{d-2^{3-\frac{4}{d}}z^{4}_{h}\rho^2(d-1)^{3-\frac{4}{d}}(d-2)^{\frac{2}{d}}[2(d-1)+(d-2)z^{4}_{h}\rho^2]^{\frac{2}{d}-2}      }
    {2^{4-\frac{4}{d}}(d-1)^{4-\frac{4}{d}}(d-2)^{\frac{2}{d}-1}[2(d-1)+(d-2)z^{4}_{h}\rho^2]^{\frac{2}{d}-2}     }      } .
\end{eqnarray}

\begin{figure}[htb]
    \centering
	\subfigure[]{
    \includegraphics[width=0.4\textwidth,height=0.6\textwidth]{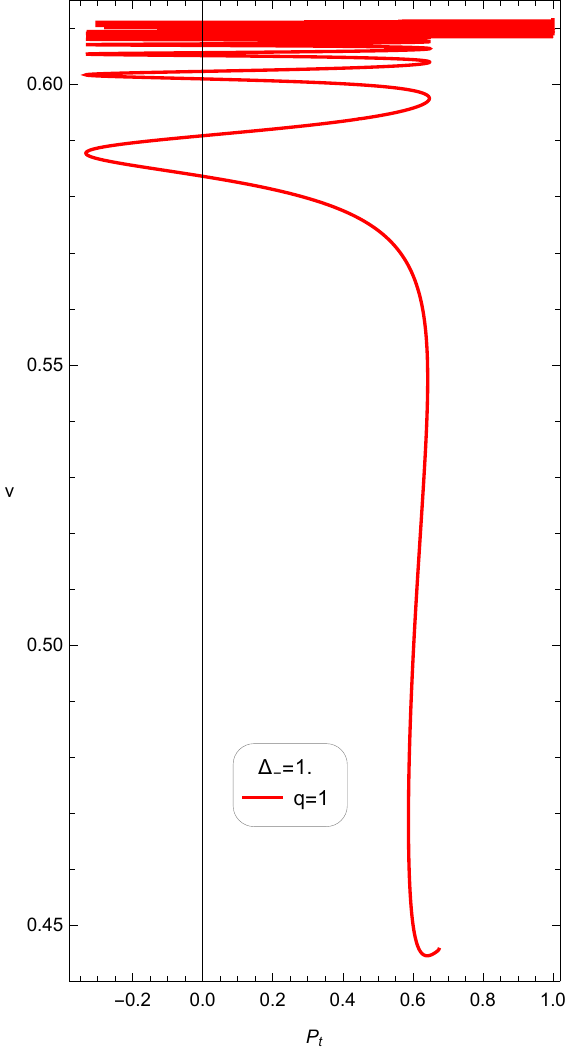}}
    \hfill
    \centering
	\subfigure[]{
	\includegraphics[width=0.4\textwidth,height=0.6\textwidth]{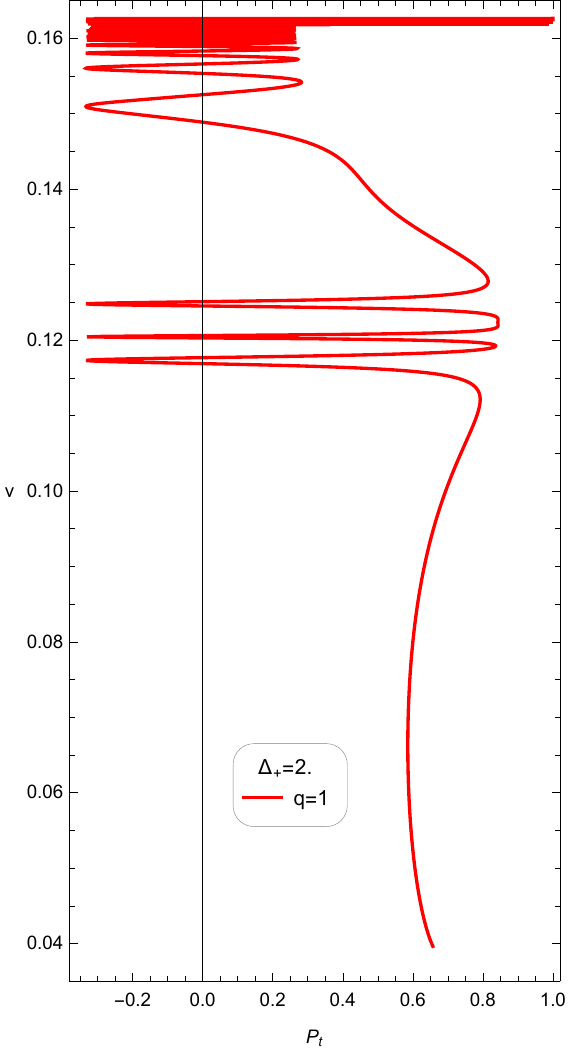}}
	\caption{The entanglement velocity $v$ as a function of the Kasner exponent $p_{t}$ 
    for the charge ($q=1$) before the Kasner inversions. }  \label{velocity1}
\end{figure}

\begin{figure}[htb]
    \centering
	\subfigure[]{
	\includegraphics[width=0.475\textwidth,height=0.35\textwidth]{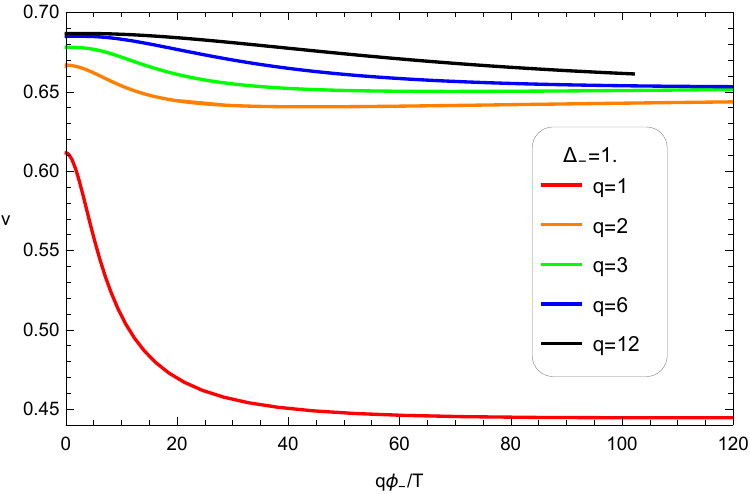}}
    \hfill
	\subfigure[]{
    \includegraphics[width=0.475\textwidth,height=0.35\textwidth]{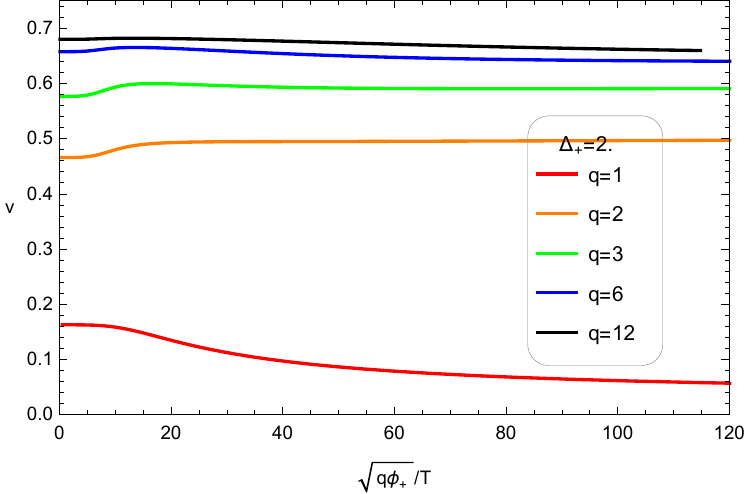}}
    \hfill\\
	\subfigure[]{
	\includegraphics[width=0.3\textwidth,height=0.221\textwidth]{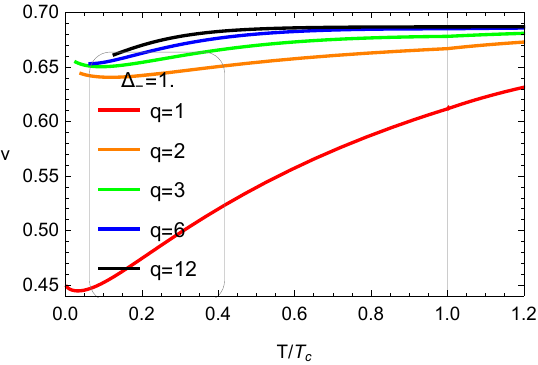}}
    \hfill
	\subfigure[]{
	\includegraphics[width=0.3\textwidth,height=0.221\textwidth]{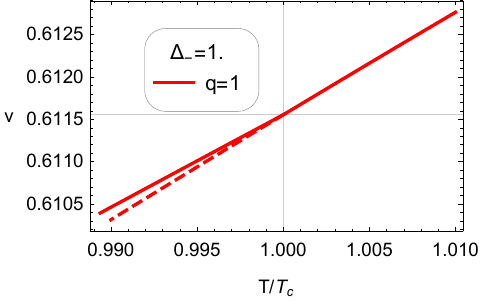}}
    \hfill
    \subfigure[]{
    \includegraphics[width=0.3\textwidth,height=0.221\textwidth]{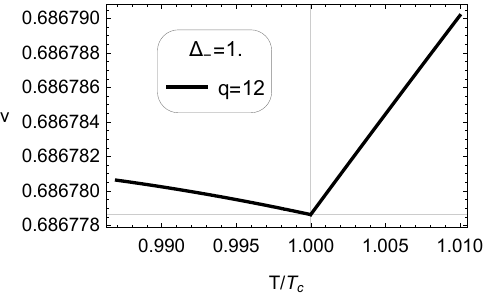}}
    \hfill\\
    \subfigure[]{
    \includegraphics[width=0.3\textwidth,height=0.221\textwidth]{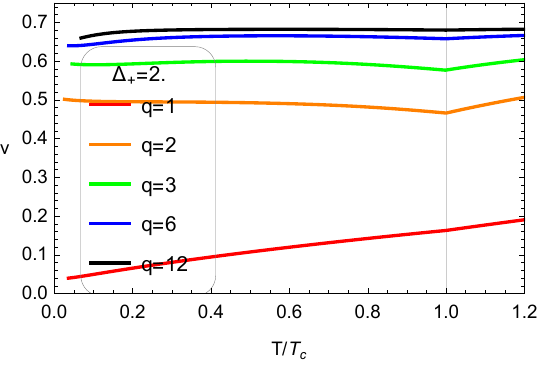}}
    \hfill
    \subfigure[]{
    \includegraphics[width=0.3\textwidth,height=0.221\textwidth]{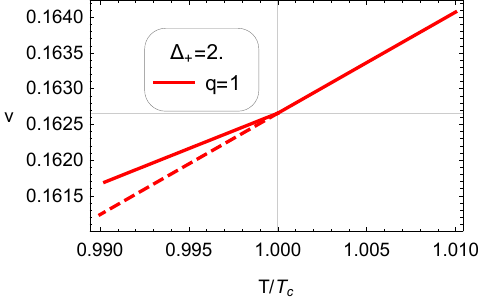}}
     \hfill
    \subfigure[]{
     \includegraphics[width=0.3\textwidth,height=0.221\textwidth]{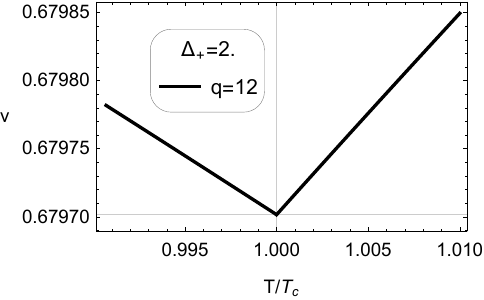}}
	\caption{ The entanglement velocity $v$ as a function of the deformation (a) $q\phi_{-}/T$; 
    (b) $\sqrt{q\phi_{+}} /T$ for different charges.   
    (c, f) represent the variation of the entanglement velocity $v$ with the temperature $T/T_{c}$,
    while (d, e, g, h) represent the situation near the critical temperature for the charge ($q=1$) and ($q=12$).}  \label{velocity11}
\end{figure}

In FIG. \ref{velocity1}, we show the entanglement velocity $v$ as a function of the Kasner exponent $p_{t}$ before the inversions. 
We find that the entanglement velocity can not only probe the interior of the black hole but also probe the Josephson oscillations,
which is significantly different from the solutions that deformed by a neutral scalar field 
\cite{Frenkel:2020ysx,Wang:2020nkd,Caceres:2021fuw,Mansoori:2021wxf}.
In FIG. \ref{velocity11}, we show the entanglement velocity $v$ as a function of 
the bath deformation $q\phi_{\pm }$ and the temperature $T/T_{c}$.
When the deformation turn off ($q\phi_{\pm }=0$), 
by comparing the critical entanglement velocity in subfigure (a, b) with the same charge,
we find that the critical entanglement velocity will change with the conformal dimension.
This behavior is different from the neutral case \cite{Caceres:2021fuw}, 
wherein the critical entanglement velocity does not change with the conformal dimension. 
This is not surprising at all. When there is no backreaction, 
our geometry is an AdS-RN black hole.
The critical temperature will change with different conformal dimensions, and  
fundamentally depends on the charge density $\rho $ of the black hole. 
In the case of neutral deformation, without backreaction, the geometry is just an AdS-Schwarzschild black hole, 
and its temperature is independent of the conformal dimension.
Similarly, the `critical' Page point, `critical' initial area difference, 
and `critical' Page time that we will discuss later all change with varying conformal dimensions.

Furthermore, we also observe that the behaviors of the entanglement velocity are related to the charges.
When the charge is small (e.g., when $q=1$), 
the entanglement velocity initially decreases with an increase in deformation (or a decrease in temperature), 
then slightly increases again as it approaches zero temperature.
It's worth noting that this behavior is more pronounced in the case of $\Delta_{-}=1$.
When the charge is large (e.g., when $q=12$), the entanglement velocity increases initially with deformation,
then decreases. This behavior is more pronounced for $\Delta_{+}=2$
(as seen in FIG. \ref{velocity11} (b, f) for the curves of $q=2$, 3, 6, and 12).
It's important to emphasize that the entanglement velocity will significantly impact on the Page time that we will investigate later.
In specific, an increase in the entanglement velocity with the deformation will lead to a decrease in the Page time, even the temperature decreases.

From subfigures (d, e, g, h) of FIG. \ref{velocity11}, 
we can observe that at the critical temperature $T_{c}$, 
the slope of the entanglement velocity shows a clear discontinuity between the superconducting ($T<T_{c}$) and normal phases ($T>T_{c}$), 
which indicates that the entanglement velocity can serve as a probe for superconducting phase transitions.
Furthermore, when we focus our attention on the difference in the entanglement velocity 
between the superconducting phase and the normal phase at the same temperature,
we find a smaller entanglement velocity difference for the chagre ($q=12$) compared to the charge ($q=1$), 
whcih is clear because the larger charge, the closer it approaches the probe limit ($q\to \infty$).

\subsection{Area of the island surface}  \label{Areaislandsurface}

As we mentioned in the introduction, 
to obtain the Page curves, the initial area of the HM surface must be smaller than the area 
of the island surface. This imposes a particular constraint on values of $x_{\mathcal{R}}$  
regardless of the temperature and conformal dimension. Specifically, the value of $x_{\mathcal{R}}$
must be larger than a critical point (the Page point) \cite{Geng:2020qvw,Ling:2020laa,Caceres:2021fuw}. 
We start from the induced metric for the time slice $(t=0)$  
\begin{eqnarray}
    ds^2 |_{t=0}=\frac{1}{z^2}\left(\frac{dz^2}{f}+dx^2_{1}+\sum_{i = 1}^{d-2}dx^{2}_{i}     \right).
\end{eqnarray}
Parameterize the coordinate $x_{1}$ as a function of $z$, such that for the any surface ($x(z)=x_{1}(z)$), the area function takes the following form
\begin{eqnarray}
    A=V_{d-2}\int \frac{dz}{z^{d-1}} \sqrt{\frac{1}{f(z)}+\dot{x}(z)^2}, \      \    \dot{x}=\frac{dx_{1}}{dz},
\end{eqnarray}
where $V_{d-2}=\int dx_2\cdots dx_{d-1} $ is an overall volume factor. 
Following the steps for calculating the area of the HM surfaces in the previous section,
we define the area density $\mathcal{A}_{I}$ and its Lagrangian $\mathcal{L}$ for the island surface
\begin{eqnarray}
    \mathcal{A}_{I}=\frac{A}{V_{d-2}}=\int \frac{dz}{z^{d-1}}\sqrt{\frac{1}{f(z) } +\dot{x}(z)^2  } =\int dz \mathcal{L}. \label{islandsurface}
\end{eqnarray}
Then, we should minimize the area density $\mathcal{A}_{I}$ and get the extremal surface that satisfies the Euler-Lagrangian equation
\begin{eqnarray}
    \frac{\partial \mathcal{L}}{\partial x}-\frac{\partial^2 \mathcal{L}}{\partial z \partial \dot{x}  }=0,
\end{eqnarray}
which indicates that $\partial \mathcal{L}/\partial \dot{x} $ is a constant of motion
\begin{eqnarray}
    \frac{\dot{x}(z)}{z^{d-1}\sqrt{\frac{1}{f(z)}+\dot{x}(z)^2 } }=C .  \label{ELeom}
\end{eqnarray}
The boundary conditions for the extremal surface are
\begin{eqnarray}
    x(0)=x_{\mathcal{R}},\      \  \frac{1}{\dot{x}(z)} \bigg |_{z=z_{T}} =0,
\end{eqnarray}
where $z_{T}$ is the turnaround point. 
This point represents the boundary of the island region on the brane. Using the Neumann condition,
the constant of motion (\ref{ELeom}) becomes
\begin{eqnarray}
    \frac{\dot{x}(z) }{z^{d-1}\sqrt{\frac{1}{f(z)}+\dot{x}(z)^2  } }=\pm \frac{1}{z^{d-1}_{T}}  
    \Rightarrow \dot{x}(z)=\pm \frac{z^{d-1}}{\sqrt{f(z)\left[z^{2(d-1)}_{T}-z^{2(d-1)}   \right]} }.   \label{eomextermal}
\end{eqnarray}
Using the Dirichlet boundary condition, we have 
\begin{eqnarray}
    x_{\mathcal{R}}=\int^{x_{\mathcal{R}}}_{0}dx=\int^{z_{T}}_{0}  \,dz \frac{z^{d-1}}{\sqrt{f(z)\left[z^{2(d-1)}_{T}-z^{2(d-1)}   \right]} }.
\end{eqnarray}
From the above equation, it is evident that given a value of $x_{\mathcal{R}}$, 
the turnaround point $z_{T}$ can be determined. 
By substituting (\ref{eomextermal}) into (\ref{islandsurface}), we can obtain the area of the island surface for a giving $x_{\mathcal{R}}$
\begin{eqnarray}
    \mathcal{A}_{I}=2\int^{z_{T}}_{0} \, \frac{dz}{z^{d-1}\sqrt{f(z)}}\frac{z^{d-1}_{T}}{\sqrt{z^{2(d-1)}_{T}-z^{2(d-1)}} }.
\end{eqnarray}
For the initial HM surface ($t=0$), 
its turnaround point $z_{T}$ does not end on the brane but approaches the singularity ($z_{T}\rightarrow \infty $).
The area expression of the initial HM surface is obtained 
by considering $\dot{x}(z)\rightarrow 0 $ as $z_{T}\rightarrow \infty $ that can be known from expression (\ref{eomextermal}).
\begin{eqnarray}
    \mathcal{A}_{HM}(t=0)=2\int^{\infty}_{0}  \, \frac{dz}{z^{d-1}\sqrt{f(z)} } =  2\int^{z_{h}}_{0}  \, \frac{dz}{z^{d-1}\sqrt{f(z)} }.
\end{eqnarray}
Because $f(z)<0$ as $z_{h}<z<\infty$, in order to avoid imaginary numbers, the integration within the black hole has been removed.    
As mentioned at the beginning of this section, there is a Page curve if and only if $\mathcal{A}_{HM}(0)<\mathcal{A}_{I}$. 
In other words, the initial area difference should satisfy
\begin{eqnarray}
    \Delta \mathcal{A}(0)=\mathcal{A}_{I}-\mathcal{A}_{HM}(0)>0.
\end{eqnarray}
Since the area of the island surface depends on $x_{\mathcal{R}}$, 
a critical Page point ($x_{p}=x_{\mathcal{R}}$) that satisfied $\Delta \mathcal{A}(0)=0$ should be found. 
Considering that the island surface is also related to the external structure of a black hole,
the initial area difference can be utilized to reflect the external backreaction.
Another important thing to note is that near the boundary ($z\rightarrow 0$), 
both of the areas are divergent due to the empty AdS$_{d+1}$ geometry
so that we will renormalize them by setting a cutoff ($\epsilon=z/z_h \ll 1$). 
The dimensionless divergent term is 
\begin{eqnarray}
    z^{d-2}_{h}\mathcal{A}_{\infty }=\frac{2}{(d-2)}\frac{1}{\epsilon^{d-2}}.
\end{eqnarray}

\begin{figure}[htb]
    \subfigure[]{
    \includegraphics[width=0.475\textwidth,height=0.33\textwidth]{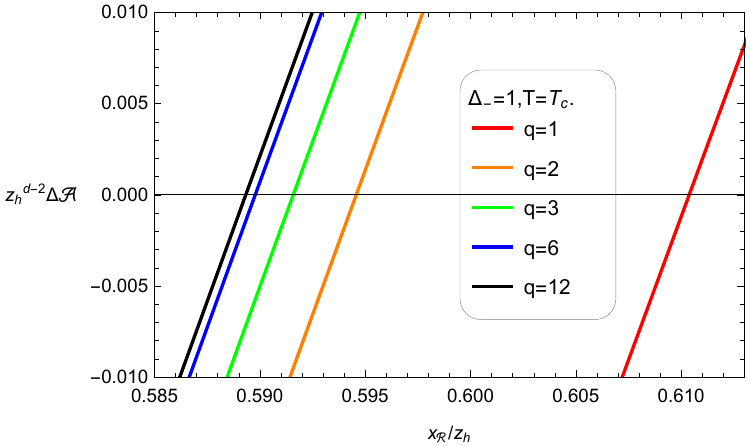}} 
    \hfill
    \subfigure[]{
    \includegraphics[width=0.475\textwidth,height=0.33\textwidth]{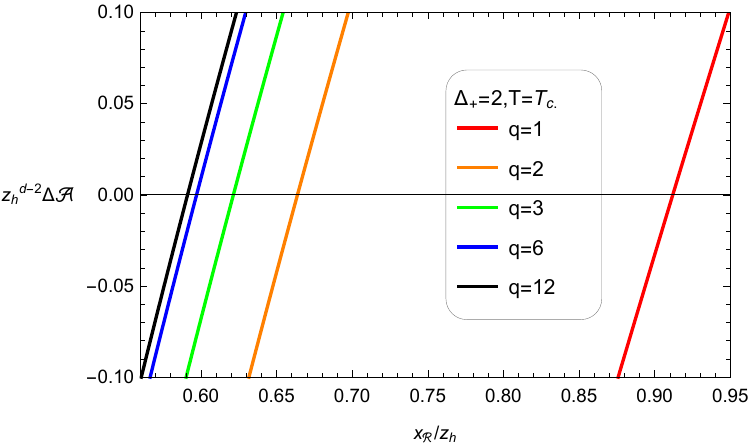}}
    \hfill\\
	\subfigure[]{
	\includegraphics[width=0.32\textwidth,height=0.23\textwidth]{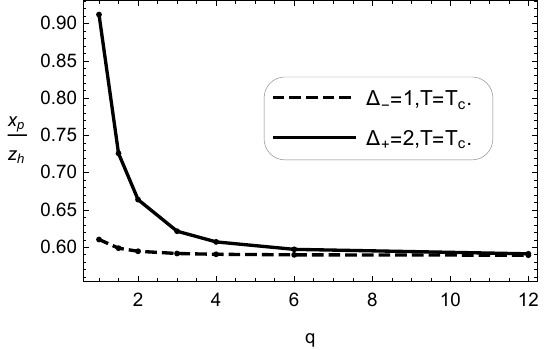}}
    \hfill
	\subfigure[]{
	\includegraphics[width=0.32\textwidth,height=0.23\textwidth]{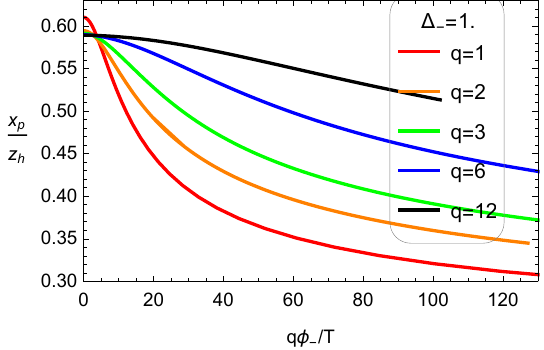}}
    \hfill
    \subfigure[]{
    \includegraphics[width=0.32\textwidth,height=0.23\textwidth]{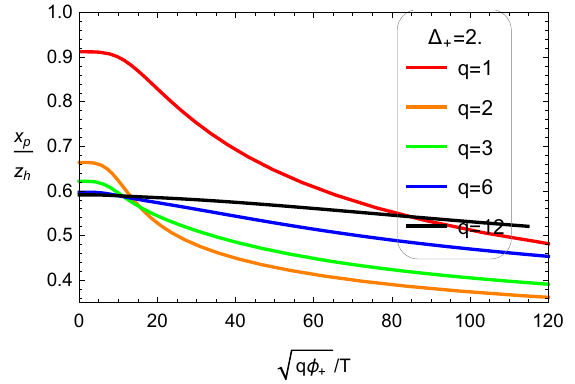}}
	\caption{The area difference $\Delta\mathcal{A}(0)$ versus $x_{\mathcal{R}}$ for different charges with 
    the conformal dimension (a) $\Delta_{-}=1$; (b) $\Delta_{+}=2$ at critical temperature $T=T_{c}$.
    (c) The Page point $x_{p}$ as a function of the charge at critical temperature.
    The Page point $x_{p}$ as a function of bath deformations (d) $q\phi_{-}/T$; (e) $\sqrt{q\phi_{+}}/T $.}  \label{pagepoint}
\end{figure}
In FIG. \ref{pagepoint}(a, b), we show the finite part of the initial area difference $\Delta \mathcal{A}(0)$  
as a function of $x_{\mathcal{R}}/z_{h}$ at the critical temperature. 
The calculation at this step is to obtain the Page point, 
which is the critical value of $x_{\mathcal{R}}$, where the area of the island surface is equal to 
the area of the initial HM surface.
From FIG. \ref{pagepoint}(c), we find that the Page point decreases as the charge increases at critical temperature,
which provides a helpful clue for us to need later to fix $x_{\mathcal{R}}$.
In FIG. \ref{pagepoint}(d, e), we also investigate the influence of scalar deformations $(q\phi_{\pm})$ on the Page point. 
We observe that the Page point decreases as the deformation increases, 
indicating that black holes require fewer radiation degrees of freedom to produce the Page curve. 
In other words, in the case of stronger deformations at the conformal boundary, 
the radiation region $\mathcal{R}$ requires more degrees of freedom to ensure the existence of the Page curve,
which provides a helpful clue for us to obtain the Page curve, 
by fixing a reasonable endpoint $x_{\mathcal{R}}$ of the radiation region $\mathcal{R}$.
Specifically, if we choose $x_{\mathcal{R}}$ larger than the critical Page point (i.e., the Page point with zero deformation $q\phi_{\pm}=0$), 
then, subsequently, no matter how much the scalar deformation increases, 
we can have a Page curve uniquely associated with the scalar deformation, 
which implies that the Page curve for a specific radiation region can probe Kasner flows \cite{Caceres:2021fuw}.

\subsection{Entanglement Entropy and Page curves}

We are now computing the Page curves for various charges and two conformal dimensions.
Based on the analysis in the previous section, we learn that to get the Page curve, the value of $x_{\mathcal{R}}$ needs to 
be larger than the Page point. Since the Page point decreases as the charge and deformation increase,
to make sure that we always have the Page curve and a finite Page time regardless of the charge and scalar deformation,
we will set $x_{\mathcal{R}}$ to be twice the critical Page point as the charge ($q=1$).
More specifically, we will fix $x_{\mathcal{R}}$
\begin{eqnarray}
     x_{\mathcal{R}}=2\frac{x_{p}}{z_h}=\left\{ \begin{aligned}   &2\times 0.610, \    \ (\Delta_{-}=1), \\
        &2\times0.912,\    \    \ (\Delta_{+}=2).
    \end{aligned} \right.   \label{2twice}
\end{eqnarray}
To obtain the Page curves, we calculate the entanglement entropy of the radiation region 
using the time-dependent HM surface that discussed in section \ref{GrowthAreaHM} before Page time. 
After the Page time, the entanglement entropy of the radiation region is provided 
by the island surface that discussed in section \ref{Areaislandsurface}. 
Since the island surface is time-independent, 
the entanglement entropy of the radiation region $\mathcal{R}$ no longer continues to increase after the Page time.
So, the Hawking radiation satisfies the unitarity of quantum mechanics.

\begin{figure}[htb] 
	\centering
	\subfigure[]{
	\includegraphics[width=0.475\textwidth,height=0.33\textwidth]{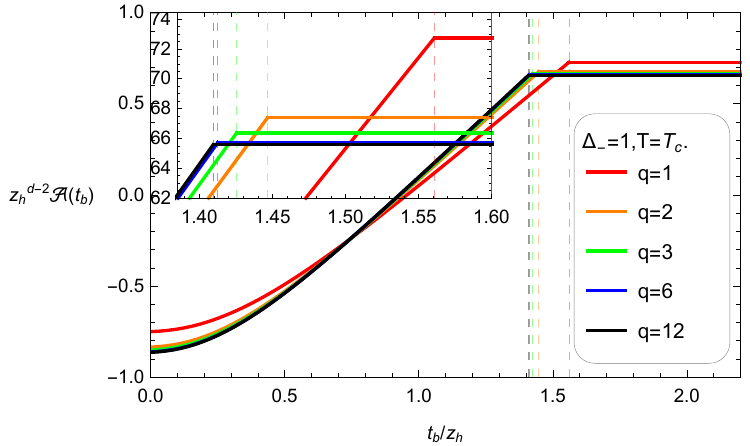}}
    \hfill
	\subfigure[]{
	\includegraphics[width=0.475\textwidth,height=0.33\textwidth]{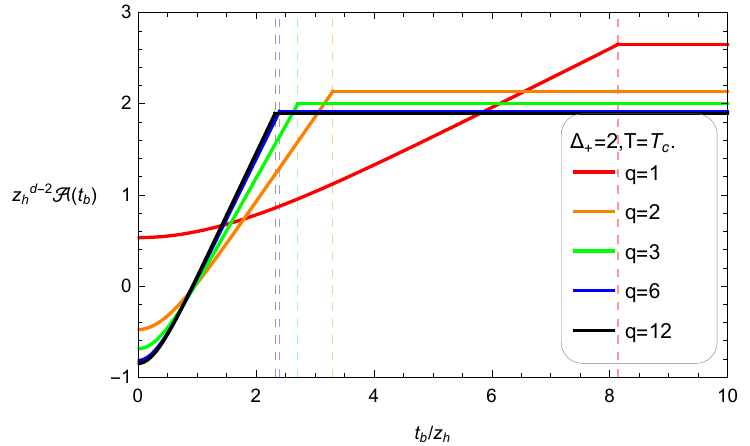}}
    \hfill\\
    \subfigure[]{
    \includegraphics[width=0.475\textwidth,height=0.33\textwidth]{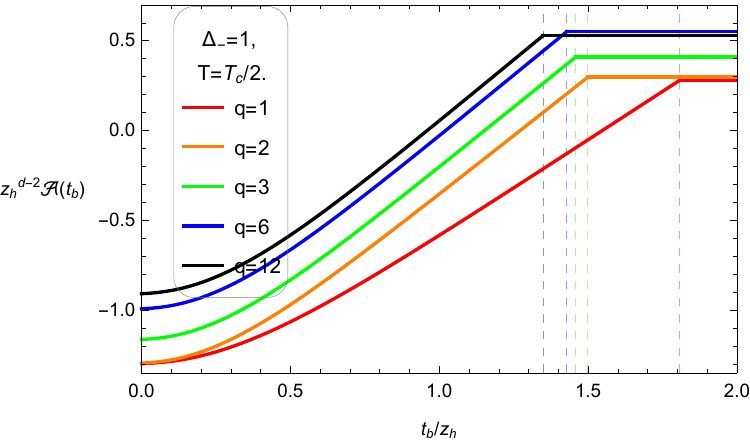}}
    \hfill
	\subfigure[]{
	\includegraphics[width=0.475\textwidth,height=0.33\textwidth]{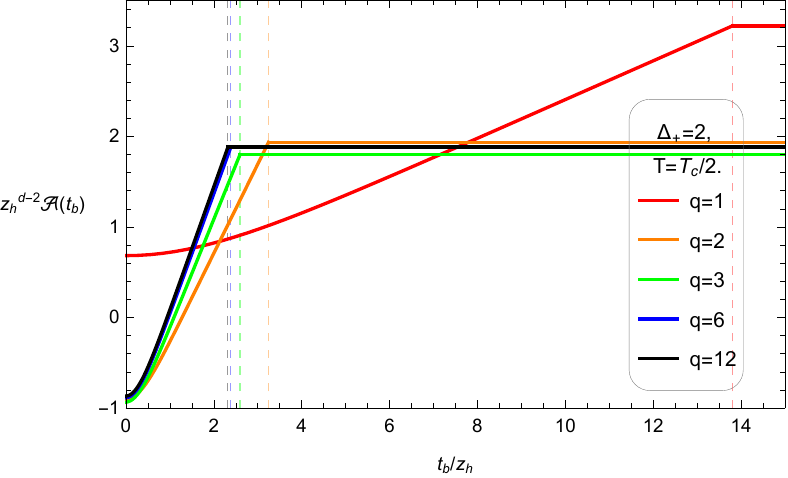}}
	\caption{The Page curves for two conformal dimension (a, c) $\Delta_{-}=1$;
    (b, d) $\Delta_{+}=2$ at the critical temperature (a, b) $T=T_{c}$ 
    and the semi-critical temperature (c, d) $T=T_{c}/2$ are
    presented in terms of the area $\mathcal{A}$ and boundary time $t_{b}$ in units of $z_{h}$.  }   \label{pagecurves}
\end{figure}  

In FIG. \ref{pagecurves}, we present the Page curves of the systems at the critical temperature (a, b) 
and semi-critical temperature (c, d) for both conformal dimensions $\Delta_{\pm }$.
From subfigures (a, b), we can observe that when the system is at the critical temperature, 
the Page time increases with the charge growth, which is because as the charge increases, 
the critical temperature of the boundary system also increases. 
According to holographic duality, the Hawking radiant temperature of the black hole increases, 
leading to a faster saturation of the radiant entanglement entropy.
However, when considering scalar deformations in the bath, 
the conventional pattern of the higher temperature leading to the shorter Page time is not necessarily valid. 
Specifically, in subfigures (c, d), we investigate the Page curves where the temperature is reduced to the half critical temperature. 
Clearly, at this point, the influence of scalar deformations and conformal dimensions becomes significant. 
By comparing subfigures (a, c), it may initially appear that as the temperature decreases, the Page time becomes longer. 
However, when we carefully examine the green curve which corresponding to $\Delta_{+}=2$ and $q=3$ in subfigures (b, d), 
we can observe that despite the temperature reduction, the Page time doesn't increase; instead, it decreases.
In FIG. \ref{pagetimes11}, we can see more intuitively that in some cases,
with increasing deformation and decreasing temperature, the Page time doesn't increase but rather decreases.
Furthermore, we also note that the entanglement velocity increases with increasing charge, 
in accordance with the results in FIG. \ref{velocity11}.

Apart from the entanglement velocity $v$, the initial area difference $\Delta\mathcal{A}(0)$ between the HM surfaces 
and island surfaces also affects the Page time. 
It raises the question of which has a more significant impact on the Page time. 
We employ a method similar to that in \cite{Caceres:2021fuw} to provide a brief discussion on this issue.
Without loss of generality, let's consider the case where the charge ($q=1$) and conformal dimension ($\Delta_{-}=1$), 
which corresponds to the red solid line in FIG. \ref{pagecurves}(a, c). 
According to our numerical calculations, the entanglement velocity, 
the initial area difference, and the Page time at critical temperature and semi-critical temperature are 
\begin{eqnarray}
    v|_{T=T_{c}}\approx 0.61155     , &&\      \  v|_{T=\frac{T_{c}}{2}}\approx 0.539485    \label{date74},\\
    \Delta \mathcal{A}(0)|_{T=T_{c}} \approx 1.474922    ,&&\      \  \Delta \mathcal{A}(0)|_{T=\frac{T_{c}}{2}} \approx 1.576144,       \label{date75}\\  
    \frac{t_{p}}{z_h}\bigg|_{T=T_{c}} \approx 1.56084         ,&&\      \   \frac{t_{p}}{z_h}\bigg|_{T=\frac{T_{c}}{2}} \approx 1.806469.      \label{date76}
\end{eqnarray}
In the assumption that the evolution of the HM surface is always linear, 
the Page time can be approximated using (\ref{envelo}) in the following manner
\begin{eqnarray}
    t_{p}\approx \frac{\Delta \mathcal{A}(0)}{2v},   
\end{eqnarray}
then we can get the ``first-order" variation in the Page time $\delta t_{p}$
\begin{eqnarray}
     \delta t_{p}\approx \frac{\delta (\Delta \mathcal{A}(0))}{2v}-\frac{\Delta \mathcal{A}(0)}{2v^2}\delta v    \label{pagetimevari}.
\end{eqnarray}
We calculate the average values of $\Delta \mathcal{A}(0)$ and $v$ by using the results  (\ref{date74}) and (\ref{date75}), 
and then substitute them into the formula above to obtain \footnote{Express the above result as the sum of two terms and setting $z_{h}=1$.}
\begin{eqnarray}
    \delta t_{p}\approx 0.0879399+0.165962=0.253902\approx 0.245629.
\end{eqnarray}
The right side of the last approximately equal sign in the above expression is from the numerical result (\ref{date76}) of the Page time.
Using the same approximate calculations, 
we can obtain the change in the Page time from the critical temperature to the semi-critical temperature 
for the case of charge ($q=1$) and conformal dimension $\Delta_{+}=2$,
\begin{eqnarray}
    \delta t_{p}\approx 1.55988+3.55321=5.11308\approx 5.65798.
\end{eqnarray}
Even these calculation are not always accurate  
\footnote{For example, in the case of $q=12$, this kind of approximate calculation yields 
results that differ significantly from the numerical results.}.
Still, it is sufficient to demonstrate that the contribution from the entanglement velocity $v$ has 
a more significant impact on the Page time than the contribution from the initial area difference $\Delta\mathcal{A}(0)$,
whcich further indicates that the backreaction from the interior of the black hole has 
a more significant influence on the Page time than the backreaction from the exterior of the black hole.
Interestingly, our results are exactly opposite to the findings regarding neutral scalar deformations in \cite{Caceres:2021fuw}.

\begin{figure}[htb]
	\centering
	\subfigure[]{
	\includegraphics[width=0.475\textwidth,height=0.35\textwidth]{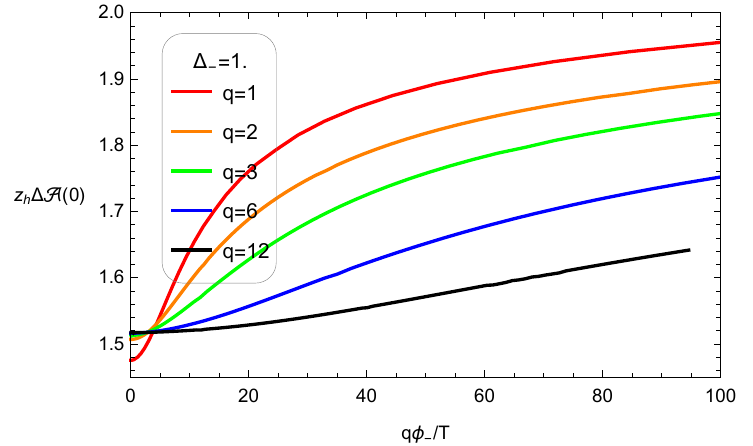}}
    \hfill
	\subfigure[]{
	\includegraphics[width=0.475\textwidth,height=0.35\textwidth]{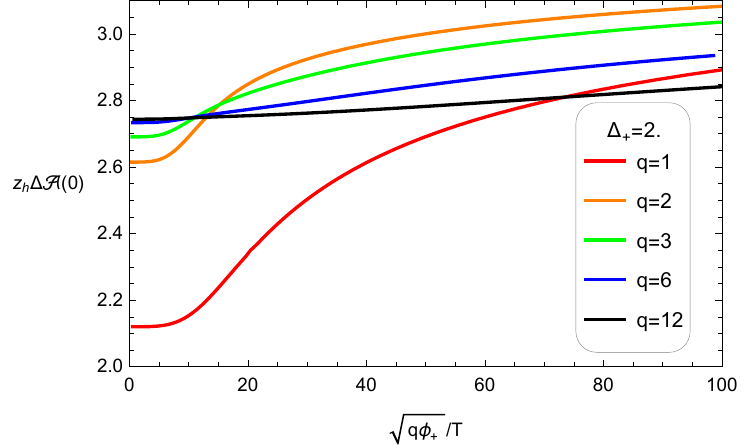}}
    \hfill\\
	\subfigure[]{
	\includegraphics[width=0.32\textwidth,height=0.221\textwidth]{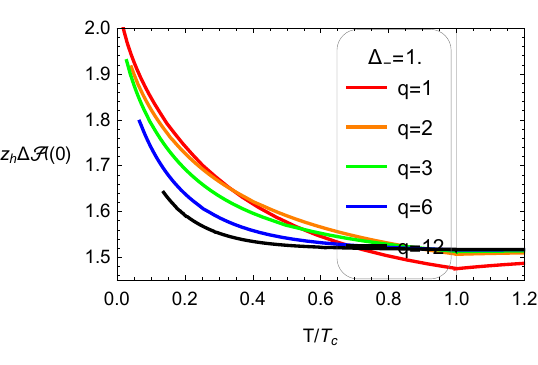}}
    \hfill
	\subfigure[]{
	\includegraphics[width=0.32\textwidth,height=0.221\textwidth]{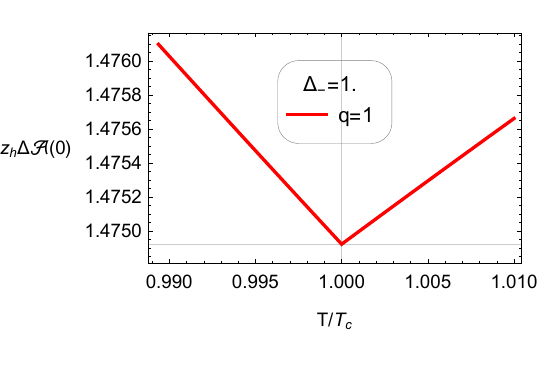}}
    \hfill
    \subfigure[]{
    \includegraphics[width=0.32\textwidth,height=0.221\textwidth]{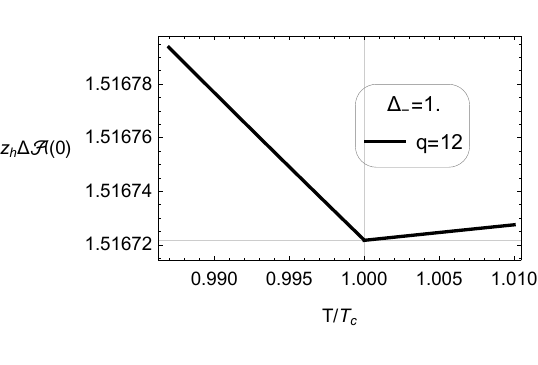}}
    \hfill\\
    \subfigure[]{
    \includegraphics[width=0.32\textwidth,height=0.221\textwidth]{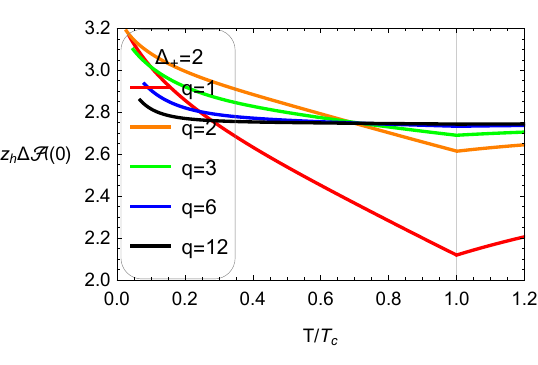}}
    \hfill
    \subfigure[]{
    \includegraphics[width=0.32\textwidth,height=0.221\textwidth]{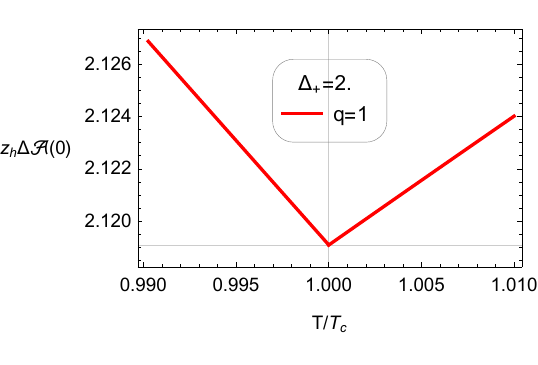}}
    \hfill
    \subfigure[]{
    \includegraphics[width=0.32\textwidth,height=0.221\textwidth]{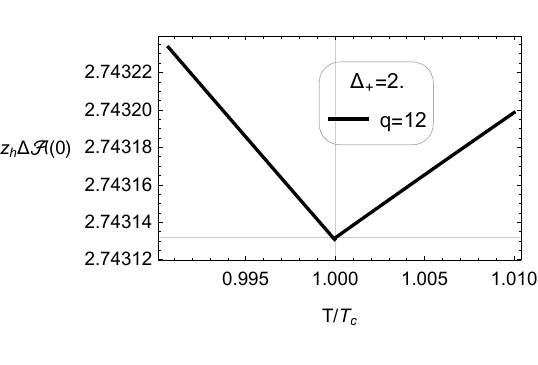}}
	\caption{The initial area differences $\Delta \mathcal{A}(0)$ are plotted as the function of 
    deformations (a) $q\phi_{-}/T $; (b) $\sqrt{q\phi_{+}} /T $.
    (c, f) represent the variation of initial area difference $\Delta \mathcal{A}(0)$ 
    with the temperature $T/T_{c}$ for the different charges,
    while (d, e, g, h) represent the situation 
    near the critical temperature for the charge ($q=1$) and ($q=12$).  }  \label{areadiff11}
\end{figure}

\begin{figure}[htb]
	\centering
	\subfigure[]{
	\includegraphics[width=0.475\textwidth,height=0.35\textwidth]{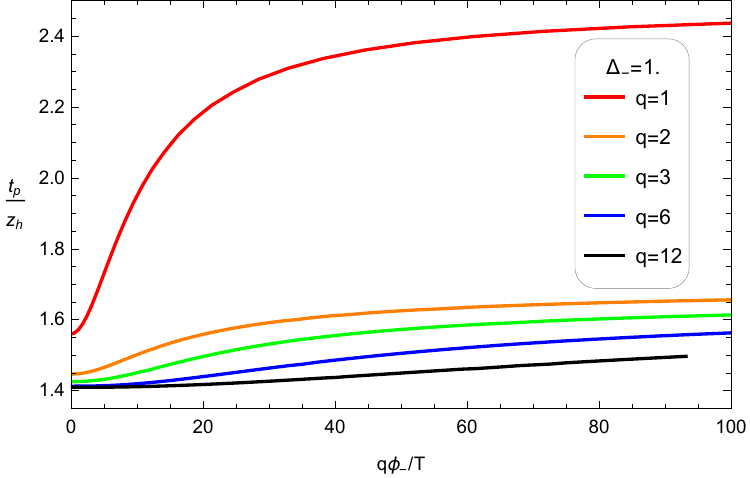}}
    \hfill
	\subfigure[]{
	\includegraphics[width=0.475\textwidth,height=0.35\textwidth]{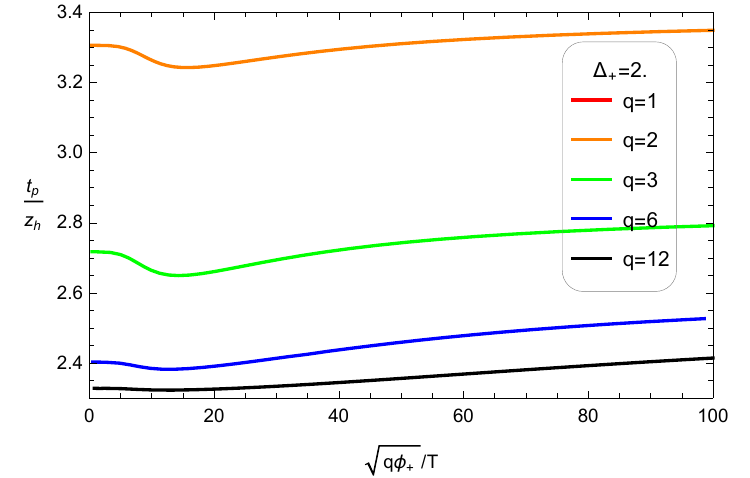}}
    \hfill\\
	\subfigure[]{
	\includegraphics[width=0.32\textwidth,height=0.221\textwidth]{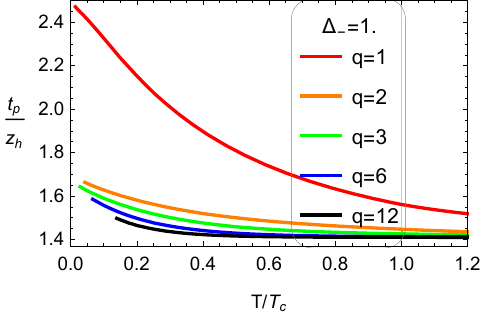}}
    \hfill
	\subfigure[]{
	\includegraphics[width=0.32\textwidth,height=0.21\textwidth]{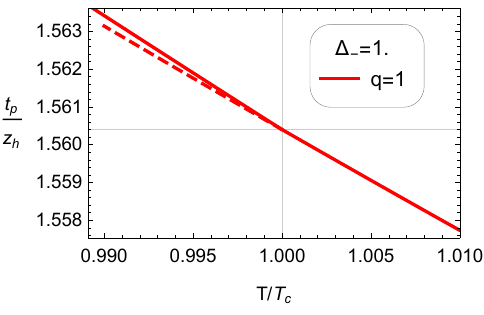}}
    \hfill
    \subfigure[]{
    \includegraphics[width=0.32\textwidth,height=0.21\textwidth]{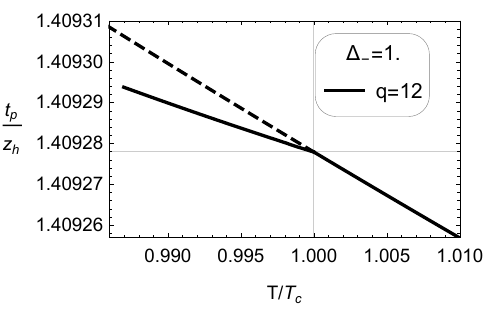}}
    \hfill\\
    \subfigure[]{
    \includegraphics[width=0.32\textwidth,height=0.221\textwidth]{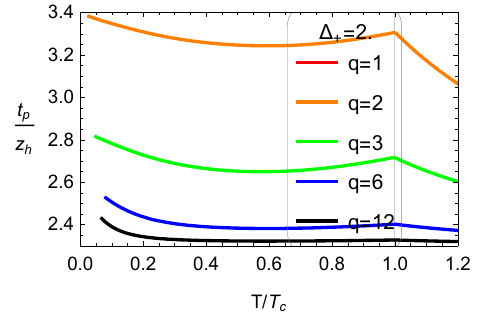}}
    \hfill
    \subfigure[]{
    \includegraphics[width=0.32\textwidth,height=0.21\textwidth]{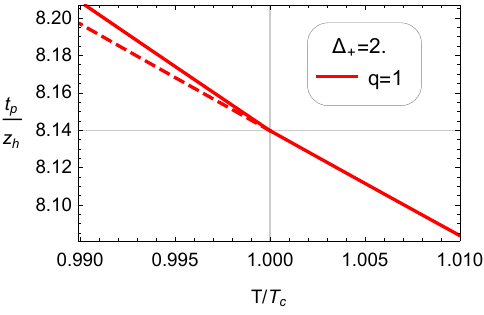}}
    \hfill
    \subfigure[]{
    \includegraphics[width=0.32\textwidth,height=0.21\textwidth]{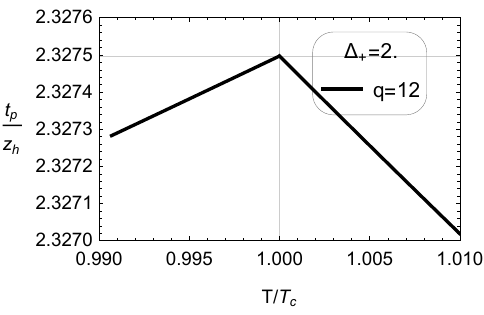}}
	\caption{The Page times $t_p$ are plotted as functions of the deformation 
	(a) $q\phi_{-}/T $; (b) $\sqrt{q\phi_{+}} /T$.
    (c, f) represent the variation of Page time $t_{p}$ with the temperature $T/T_{c}$ for different charges,
    while (d, e, g, h) represent the situation near the critical templerature for the chagre $q=1$ and $q=12$.
    Since the page time for the case ($\Delta_{+}=2$, $q=1$) is too big compared with other cases, 
    it is not convenient to be presented in the same figure (b). }  \label{pagetimes11}
\end{figure}

In FIG. \ref{areadiff11} and \ref{pagetimes11},
we provide the initial area difference and Page time as a function of bath deformation and temperature, respectively. 
We found that the initial area difference increases with increasing scalar deformation (decreasing temperature). 
Since greater deformation leads to a smaller Page point, keeping the value of $x_{\mathcal{R}}$ constant results in an 
increase in the initial area difference as the Page point becomes smaller. 
The similar curves between FIG. \ref{areadiff11}(a, b) and FIG. \ref{pagepoint}(d, e) (except one increasing and one decreasing)
indicates a connection between the initial area difference and the Page point.
We also observe that the critical initial area difference (when $q\phi_{\pm}=0$) increases with an increase in the charge. 
This is because at the critical temperature, the Page point decreases as the charge increases, as seen in the FIG. \ref{pagepoint}(c). 
Due to the clear discontinuity in the slope of the initial area difference at the critical temperature, 
the initial area difference in the Page curve can serve as a valuable probe for superconducting phase transitions.

From the curves of all charges in FIG. \ref{pagetimes11}(a) and the curves for charge ($q=1$) in FIG. \ref{pagetimes11}(b),
it is not difficult to observe that the Page time increases with the scalar deformation.
However, for the cases of the conformal dimension ($\Delta_{+}=2$) and charges ($q=2$, $3$, $6$, and $12$) in FIG. \ref{pagetimes11}(b), 
the Page time initially decreases and then increases with deformation.
This behavior is related to the corresponding behavior of the entanglement velocity, 
which increases initially and then decreases with deformation in FIG. \ref{velocity11}(b).
This further indirectly validates our earlier conclusion that the entanglement velocity 
has a more significant impact on Page time than the initial area difference.
In other words, the internal backreaction has a more significant impact 
on Page time than external backreaction.
From FIG. \ref{pagetimes11} (d, e, g, h), 
we can observe that the slope of the Page time with respect to temperature 
is discontinuous at the critical temperature. 
This indicates that the Page time can serve as a valuable probe 
for detecting superconducting phase transitions.

It is worth noting that in all calculations above, 
we have fixed the value of $x_{\mathcal{R}}$ to be twice the critical Page point (\ref{2twice}).
In fact, whether internal or external backreaction has a greater impact on the Page time 
also depends on the value of $x_{\mathcal{R}}$, which is related to the radiation region $\mathcal{R}$ \cite{Caceres:2021fuw}.
To study this issue, we consider the variation in the initial area difference $\delta(\Delta \mathcal{A}(0))$ 
and the variation in the Page time $\delta t_{p}$ between the critical temperature $T=T_{c}$ and semi-critical temperature $T=T_{c}/2$. 
Without loss of generality, we only consider the case of charge ($q=1$),
while keeping the other constants set as $d=3$ and $m^2=-2$.
More specifically, we consider the following variation in the initial area difference and the variation in the Page time
\begin{eqnarray}
    \delta(\Delta \mathcal{A}(0))&=&\Delta \mathcal{A}(0)|_{T=T_{c}/2}- \Delta \mathcal{A}(0)|_{T=T_{c}},\\
               \delta t_{p}&=&t_{p}|_{T=T_{c}/2}-t_{p}|_{T=T_{c}}.
\end{eqnarray}

\begin{figure}[htb] 
	\centering
	\subfigure[]{
	\includegraphics[width=0.475\textwidth,height=0.33\textwidth]{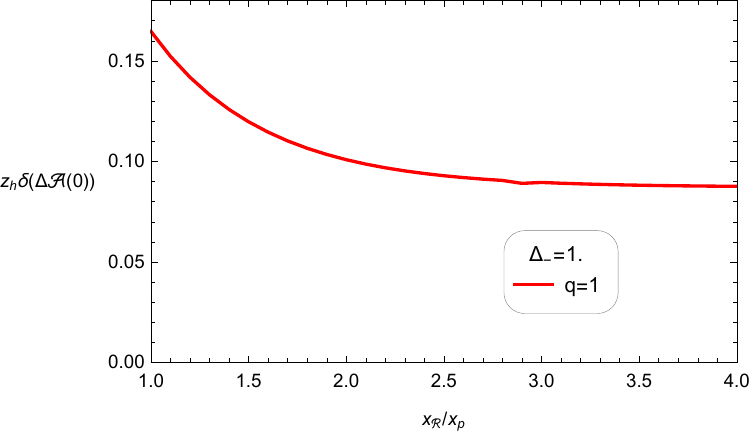}}
    \hfill
	\subfigure[]{
	\includegraphics[width=0.475\textwidth,height=0.33\textwidth]{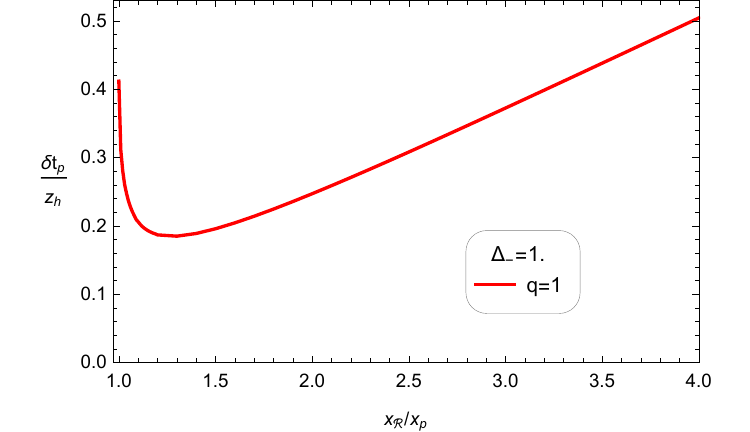}}
    \hfill\\
    \subfigure[]{
    \includegraphics[width=0.475\textwidth,height=0.33\textwidth]{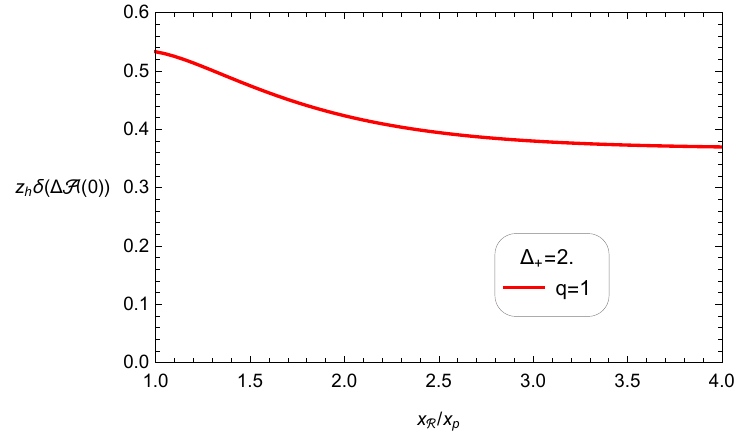}}
    \hfill
	\subfigure[]{
	\includegraphics[width=0.475\textwidth,height=0.33\textwidth]{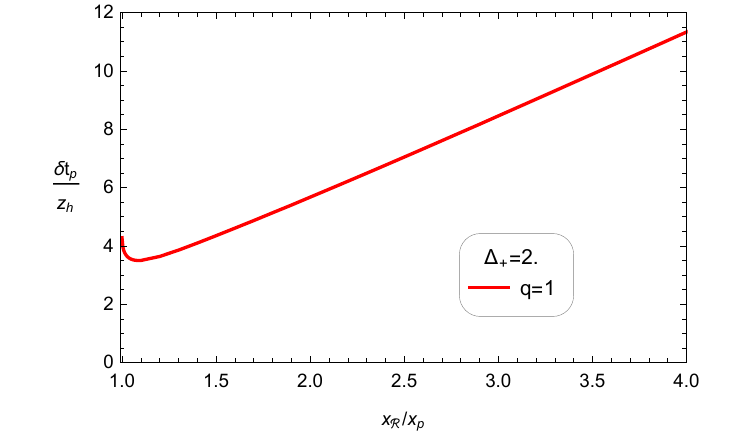}}
	\caption{The variations in the initial area difference and the Page time between 
    the critical temperature and semi-critical temperature are
     presented in terms of radition $x_{\mathcal{R}}$ region in units of $z_{h}$ .  }   \label{variationfig}
\end{figure}

In FIG. \ref{variationfig}, 
the variations in the initial area difference and the Page time are
presented in terms of $x_{\mathcal{R}}$. 
From FIG. \ref{variationfig}, it is evident that the variation in the initial area difference decreases as $x_{\mathcal{R}}$ increases.
However, according to (\ref{pagetimevari}), the variation in Page time should also decrease continuously.
In fact, the variation in Page time decreases initially with increasing $x_{\mathcal{R}}$ and then linearly increases.
Hence, we can conclude that the initial decrease (when $x_{\mathcal{R}}\approx x_{p}$) 
in the variation of the Page time is contributed by the variation in the initial area difference.
The subsequent linear increase (when $x_{\mathcal{R}} \gg x_{p}$) in the variation of the Page time 
is attributed to the second term in expression (\ref{pagetimevari}),
which is the contribution of the variation in the entanglement velocity.
In other words, when $x_{\mathcal{R}}\approx x_{p}$, the external backreaction has a significant impact on the Page time. 
When $x_{\mathcal{R}}\gg x_{p}$, the internal backreaction has a more substantial influence on the Page time.
Our conclusion aligns perfectly with the findings regarding neutral scalar deformations \cite{Caceres:2021fuw}.
Finally, it is worth mentioning that in our model, $x_{\mathcal{R}}=2x_{p}$ falls within the $x_{\mathcal{R}}\gg x_p$ region,
which is why the previous calculations conclude that the entanglement velocity has a more significant impact on Page time than the initial area difference.
However, in the case of neutral scalar deformations \cite{Caceres:2021fuw}, $x_{\mathcal{R}}=2x_{p}$ falls within the $x_{\mathcal{R}}\approx x_{p}$ region,
and that's why the conclusion is drawn that external backreaction has a more significant influence on Page time than internal backreaction.

\section{Conclusions and Outlook}    \label{section06}
In this paper, we employ a dual holographic model that deforms the conformal bath 
with a charged scalar field to investigate how the bath deformation affects the Page curve. 
Unlike the neutral deformation, due to the spontaneous symmetry breaking of the U(1) gauge field, 
it induces a second-order phase transition of the charged scalar field, known as the holographic superconductors.

We have discovered that the characteristic parameters of the Page curve, such as entanglement velocity,
initial area difference and Page time, can be used as suitable probes to detect the superconducting phase transitions. 
Specifically, the entanglement velocity can probe both the Kasner flows and Josephson oscillations.

When keeping  $x_{\mathcal{R}}$ fixed, 
we find that the growth behavior of the Page time with an increasing deformation parameter
(or decreasing temperature) is related to the conformal dimension $\Delta_{\pm}$.
Specifically, the Page time monotonically increases with an increasing deformation for $\Delta_{-}=1$ 
(corresponding to the ``Neumann'' ($\Delta=\Delta_{-}$) quantization).
However, for $\Delta_{+}=2$ 
(corresponding to the ``Dirichlet'' ($\Delta=\Delta_{+}$) quantization),
and charges ($q=2$, $3$, $6$, and $12$) the Page time initially decreases and then increases with an increasing deformation (or a decreasing temperature).
The behavior of the Page time decreasing as the temperature lowers is closely related to the corresponding behavior of the entanglement velocity.
In the case of charge ($q=1$), the behavior of the Page time is similar to the former.
For $\Delta_{+}$, the Page time is generally larger than the Page time for $\Delta_{-}$. 
Furthermore, a larger charge leads to a smaller Page time for the same conformal dimension and fixed temperature.

When the fixed $x_{\mathcal{R}}$ 
equals twice the critical Page point $x_{\mathcal{R}}=2x_{p}$, the change in the entanglement velocity (the internal backreaction) 
has a more significant influence on the Page time compared to the change in the initial area difference (the external backreaction). 
By varying the value of $x_{\mathcal{R}}$, we further discover that when $x_{\mathcal{R}}\gg x_{p}$, 
meaning the black hole has more boundary degrees of freedom, 
the internal backreaction has a more significant effect on Page time compared to the external backreaction,
which is consistent with results in the case of neutral scalar deformations \cite{Caceres:2021fuw}.

In recent years, an information-theoretic quantity known as computational complexity 
has been related to the gravitational concept within the context of AdS/CFT duality \cite{Susskind:2014rva,Stanford:2014jda,Brown:2015bva,Brown:2015lvg,Alishahiha:2015rta}.
Interestingly, the holographic complexity 
has also received extensive research in the holographic superconducting model \cite{KordZangeneh:2017zyy,Fujita:2018xkl,Xu:2022qjm,Wang:2023nes,Yang:2019gce,An:2022lvo,Auzzi:2022bfd}.
Inspired by previous researchs in the holographic complexity with the doubly holographic model \cite{Bhattacharya:2020uun,Hernandez:2020nem,Bhattacharya:2021jrn,Bhattacharya:2021jrn,Bhattacharya:2021nqj}.
In the future, we're interested in exploring the holographic complexity of the holographic superconducting models but with a doubly holographic setup.

\acknowledgments
YC Xu would like to thank Prof. Qiu Taotao of Huazhong University of Science and Technology 
for his helpful discussions and suggestions.
This work was supported by the National Natural
Science Foundation of China under Grants No. 11653002 and No. 11875141,

\bibliography{references}
\bibliographystyle{apsrev4-1}

\end{document}